\RequirePackage[l2tabu, orthodox]{nag}
\RequirePackage{snapshot}

\documentclass[9pt,twocolumn]{extarticle}

\makeatletter
\if@twocolumn
  \usepackage[dvips,letterpaper,top=0.5in, bottom=0.5in, left=0.75in, right=0.5in,includefoot,heightrounded]{geometry}
\else
  \usepackage[dvips,letterpaper,margin=1in,includefoot,heightrounded]{geometry}
\fi

\usepackage{srcltx}

\usepackage{amsmath}
\usepackage{amssymb,amsfonts}

\usepackage{abstract}

\usepackage[usenames,dvipsnames]{color}
\usepackage[usenames,dvipsnames]{xcolor}

\usepackage{setspace}

\usepackage{url}

\usepackage{mwe}

\usepackage{subfigure}
\usepackage{graphicx}
\usepackage[small]{caption}
\usepackage{multicol}
\usepackage{multirow}

\usepackage{amsmath}
\usepackage{amssymb,amsfonts}
\usepackage{textcomp}

\usepackage{algorithm}
\usepackage{algpseudocode}

\usepackage{booktabs}
\usepackage{cite}

\usepackage{flushend}
\usepackage{bm}
\usepackage[normalem]{ulem}
\usepackage{mathptmx}
\usepackage[english]{babel}

\usepackage[short,12hr]{datetime}

\usepackage{hyperref}
\usepackage[hyperpageref]{backref}

\makeatletter

\newcommand{\printtitle}{%
\makeatletter
\if@twocolumn

\twocolumn[%
  \maketitle
  \begin{onecolabstract}
    \myabstract
  \end{onecolabstract}
  \begin{center}
    \small
    \textbf{Keywords}
    \\\medskip
    \mykeywords
  \end{center}
  \bigskip
]
\saythanks
\else
  \maketitle
  \begin{onecolabstract}
    \myabstract
  \end{onecolabstract}
  \begin{center}
    \small
    \textbf{Keywords}
    \\\medskip
    \mykeywords
  \end{center}
  \bigskip
  \onehalfspacing
\fi
\makeatother
}

\title{%
Low-complexity 8-point DCT Approximation
Based on Angle Similarity
for Image and Video Coding}

\author{Ra\'iza~S.~Oliveira
\thanks{R. S. Oliveira
is with the
Programa de P\'os-Gradua\c{c}\~ao em Engenharia El\'etrica,
Universidade Federal de Pernambuco (UFPE), Recife,
Brazil;
and with the
Signal Processing Group,
Departamento de Estat\'{\i}stica,
UFPE.}
\and
Renato~J.~Cintra
\thanks{%
Renato~J.~Cintra
is with the
Signal Processing Group,
Departamento de Estat\'{\i}stica,
Universidade Federal de Pernambuco,
Recife, Brazil;
ECE,
University of Calgary,
Calgary, AB, Canada.
E-mail: \mbox{rjdsc@de.ufpe.br}
}
\and
F\'abio~M.~Bayer
\thanks{F. M. Bayer
is with
the
Departamento de Estat\'{\i}stica,
Universidade Federal de Santa Maria,
Santa Maria,
and
LACESM,
Brazil.}
\and
Thiago~L.~T.~da~Silveira
\thanks{T.~L.~T.~Silveira
is with the Programa de P\'os-Gradua\c{c}\~ao em Computa\c{c}\~ao,
Universidade Federal do Rio Grande do Sul, Porto Alegre,
Brazil.}
\and
Arjuna~Madanayake%
\thanks{%
Arjuna Madanayake
is with the
Department of Electrical and Computer Engineering,
University of Akron, Akron, OH.}
\and
Andr\'e Leite
\thanks{%
Andr\'e Leite
is with the
Departamento de Estat\'{\i}stica,
Universidade Federal de Pernambuco,
Recife, Brazil.
E-mail: \mbox{leite@de.ufpe.br}
}
}

\date{}

\newcommand{\myabstract}{%
The principal component analysis (PCA)
 is widely used for data decorrelation
 and dimensionality reduction.
 However,
 the use of PCA may be impractical in real-time applications,
 or in situations were energy and computing constraints
are severe.
 In this context,
 the discrete cosine transform (DCT)
 becomes a low-cost alternative
 to data decorrelation.
 This paper presents a method
 to derive
 computationally efficient approximations
 to the DCT.
 The proposed method
 aims at the minimization of
 the angle
 between the rows of the exact DCT matrix
 and the rows of the approximated transformation matrix.
 The resulting transformations matrices
 are orthogonal
 and have extremely low arithmetic complexity.
 Considering popular performance measures,
 one of the proposed transformation matrices
 outperforms the best competitors
 in both matrix error and coding capabilities.
Practical applications
in image and video coding
demonstrate the relevance
of the proposed transformation.
In fact,
we show that
the proposed approximate DCT
can outperform
the exact DCT for image encoding
under certain compression ratios.
The proposed transform
and its direct competitors
are also physically realized
as digital prototype circuits
using FPGA technology.
}

\newcommand{\mykeywords}{%
DCT Approximation,
Fast algorithms,
Image/video encoding
}

\begin{document}

\printtitle

\section{Introduction} \label{s:introduction}

Data decorrelation
is a central task in many statistical and signal processing
problems~\cite{dunteman1989pca,jolliffe2002pca,britanak2007discrete}.
Decorrelation
can be accomplished
by means of a linear transformation
that converts correlated observations
into linearly uncorrelated values.
This operation is commonly performed by
principal component analysis~(PCA)~\cite{jolliffe2002pca}.
PCA is widely used
to reduce the dimensionality of data~\cite{jolliffe2002pca, Gorban2007},
where
the information contained in all the original variables
is replaced by
the
data variability information
of the initial few uncorrelated principal components.
The quality of such approximation
depends on the number of components used
and the proportion of variance explained, or energy retained, by each of them.

In
the field of analysis and processing of images and signals,
PCA, also known
as the discrete Karhunen--Lo\`eve transform (KLT)~\cite{britanak2007discrete},
is considered the optimal
linear transformation
for data decorrelation
when the signal is a first order Markov process~\cite{rao1990,britanak2007discrete}.
The KLT has the following features~\cite{britanak2007discrete}:
(i)~decorrelates the input data completely in the transform domain;
(ii)~minimizes the mean square error in data compression;
and
(iii)~concentrates the energy (variance) in a few coefficients of the output vector.
Because
the KLT matrix
depends on the variance and covariance matrix of the input data,
deriving
computationally efficient
algorithms for real-time processing
becomes a very hard task~\cite{chen2012, alvarez2017, bae2017, thomakos2016, zeng2017, blahut2010, ahmed1974, britanak2007discrete, Clarke1981}.

If
the
input data follows
a
stationary
highly correlated
first-order
Markov process~\cite{ahmed1974, britanak2007discrete, Cintra2014},
then
the KLT is
very closely approximated by
the discrete cosine transform (DCT)~\cite{ahmed1974, britanak2007discrete}.
Natural images fall into this particular statistical model category~\cite{Hall2006}.
Thus DCT
inherits the  decorrelation and compaction
properties of the KLT,
with the advantage of having a closed-form expression
independent of the input signal.
Image and video communities
widely adopt the DCT in
their most successful
compression standards,
such as JPEG~\cite{Wallace1992jpeg} and MPEG~\cite{Puri2004}.
Often such standards include
two-dimensional (\mbox{2-D})
versions of the DCT
applied
to small image blocks
ranging
from
4$\times$4
to
32$\times$32
pixels.

The 8$\times$8
block is
employed in a large number of
standards,
for example:
JPEG~\cite{Wallace1992jpeg},
MPEG~\cite{le1992mpeg},
H.261~\cite{h261},
H.263~\cite{h263},
H.264/AVC~\cite{h2642003},
and
HEVC~\cite{pourazad2012}.
The arithmetic cost
of the 8-point DCT
is 64~multiplications and 56~additions,
when computed by definition.
Fast algorithms
can dramatically reduce the arithmetic cost
to
11~multiplications and 29~additions,
as in the Loeffler DCT algorithm~\cite{loeffler1991}.

However,
the number of DCT calls
in
common image and video
encoders is extraordinarily high.
For instance,
a single image frame of high-definition TV (HDTV)
contains 32.400
8$\times$8 image subblocks.
Therefore,
computational savings
in the DCT step
may effect
significant performance gains,
both in terms
of
speed
and
power consumption~\cite{sadhvi2014,coutinho2015}.
Being quite a mature area of research~\cite{bayer201216pt},
there is little room for improvement
on
the exact DCT computation.
Thus,
one approach
to further minimize
the computational cost
of computing the DCT
is the use of matrix approximations~\cite{Cintra2010,Cintra2014}.
Such approximations
provide matrices with
similar mathematical behavior
to the exact DCT
while
presenting
a
dramatically
low arithmetic cost.

The goals of this paper are as follows.
First,
we aim
at establishing
an optimization problem
to facilitate the derivation
of 8-point DCT approximations.
To this end,
we adopt a
vector angle based objective function
to minimize the angle
between the rows of the approximate
and the exact DCT matrices
subject to
orthogonality constraints.
Second,
the sought approximations are
(i)~submitted
to a comprehensive assessment
based on well-known figures of merit
and
(ii)~compared to state-of-the-art
DCT approximations.
Third,
fast algorithms are derived
and
realized
in FPGA hardware
with
comparisons with competing methods.
We also examine the performance
of the obtained transformations
in the context of image compression
and
video coding.
We demonstrate
that
one of our DCT approximations
can outperform
the exact DCT
in terms of
effected quality
after image compression.

This paper is organized as follows.
In Section~\ref{s:dctapp},
the 8-point DCT
and
popular DCT approximations
are discussed.
Section~\ref{s:method}
introduces an optimization problem
to pave the way for the derivation
of new DCT approximations.
In Section~\ref{s:measures}
the proposed approximations
are detailed
and
assessed
according
to well-known performance measures.
In Section~\ref{s:hardware}
a fast algorithm
for the proposed approximation
is presented.
Moreover,
a field-programmable gate array~(FPGA)
design is proposed
and compared with competing methods.
Section~\ref{s:aplication}
furnishes computational evidence
of the appropriateness
of the introduced approximate DCT
for image and video encoding.
Section~\ref{s:conclusion}
concludes the paper.

\section{DCT Approximations} \label{s:dctapp}

Let $\mathbf{x}$ and $\mathbf{X}$ be 8-point column vectors
related by the DCT.
Therefore,
they satisfy the following expression:
$\mathbf{X} = \mathbf{C} \cdot \mathbf{x}$,
where
\begin{align*}
\mathbf{C} =
\left[
\begin{smallmatrix}
\gamma_3  &  \phantom{-}\gamma_3  &  \phantom{-}\gamma_3  &  \phantom{-}\gamma_3  &  \phantom{-}\gamma_3  &  \phantom{-}\gamma_3  &  \phantom{-}\gamma_3  &  \phantom{-}\gamma_3\\
\gamma_0  &  \phantom{-}\gamma_2  &  \phantom{-}\gamma_4  &  \phantom{-}\gamma_6  & -\gamma_6             & -\gamma_4             & -\gamma_2             & -\gamma_0\\
\gamma_1  &  \phantom{-}\gamma_5  & -\gamma_5             & -\gamma_1             & -\gamma_1             & -\gamma_5             &  \phantom{-}\gamma_5  &  \phantom{-}\gamma_1\\
\gamma_2  & -\gamma_6             & -\gamma_0             & -\gamma_4             & \phantom{-} \gamma_4  &  \phantom{-}\gamma_0  &  \phantom{-}\gamma_6  & -\gamma_2\\
\gamma_3  & -\gamma_3             & -\gamma_3             &  \phantom{-}\gamma_3  &  \phantom{-}\gamma_3  & -\gamma_3             & -\gamma_3             &  \phantom{-}\gamma_3\\
\gamma_4  & -\gamma_0             &  \phantom{-}\gamma_6  &  \phantom{-}\gamma_2  & -\gamma_2             & -\gamma_6             &  \phantom{-}\gamma_0  & -\gamma_4\\
\gamma_5  & -\gamma_1             &  \phantom{-}\gamma_1  & -\gamma_5             & -\gamma_5             &  \phantom{-}\gamma_1  & -\gamma_1             &  \phantom{-}\gamma_5\\
\gamma_6  & -\gamma_4             &  \phantom{-}\gamma_2  & -\gamma_0             & \phantom{-} \gamma_0  & -\gamma_2             &  \phantom{-}\gamma_4  & -\gamma_6\\
\end{smallmatrix}
\right],
\end{align*}
and
$\gamma_k = \cos(2{\pi}(k+1)/32)$,
for $k= 0, 1, \ldots, 6$.

Common algorithms
for efficient DCT computation
include:
(i)~Yuan \emph{et~al.}~\cite{Yuan2006},
(ii)~Arai \emph{et~al.}~\cite{arai1988fast},
(iii)~Chen \emph{et~al.}~\cite{Chen1977},
(iv)~Feig--Winograd~\cite{fw1992},
(v)~Lee~\cite{lee84},
and
(vi)~Hou~\cite{hou87}.
Table~\ref{tab:dctfat}
lists the computational costs
associated to such methods.
The theoretical minimal
multiplicative complexity is 11~multiplications~\cite{Heideman1988,loeffler1991},
which
is attained
by
the Loeffler algorithm~\cite{loeffler1991}.

\begin{table}
\centering
\caption{Computational cost of the fast algorithms for the DCT}
\label{tab:dctfat}

\begin{tabular}{lcc}
\toprule
Algorithm   &  Multiplications & Additions \\
\midrule
Loeffler \emph{et~al.}~\cite{loeffler1991, Liang2001, masera2017odd} & 11 & 29 \\
Yuan \emph{et~al.}~\cite{Yuan2006, wang2011combined}         & 12 & 29 \\
Lee~\cite{lee84, snigdha2016, suzuki2010integer}                           & 12 & 29 \\
Hou~\cite{hou87, fong2012, choi2010}                           & 12 & 29 \\
Arai \emph{et~al.}~\cite{arai1988fast, Liang2001, masera2017}              & 13 & 29 \\
Chen \emph{et~al.}~\cite{Chen1977, park2012, Liang2001}         & 16 & 26 \\
Feig--Winograd~\cite{fw1992, xu2016}              & 22 & 28 \\
\bottomrule
\end{tabular}
\end{table}

A DCT approximation
is a matrix~$\widehat{\mathbf{C}}$
capable of furnishing
$
\widehat{\mathbf{X}}
=
\widehat{\mathbf{C}}
\cdot
\mathbf{x}
$
where
$
\widehat{\mathbf{X}}
\approx
\mathbf{X}
$
according to some
prescribed criterion,
such as
matrix proximity
or
coding performance~\cite{britanak2007discrete}.
In general terms,
as shown in~\cite{Higham1986, cintra2011dct, cintra2011, britanak2007discrete, Cintra2012},
$\widehat{\mathbf{C}}$ is a real valued matrix which
consists
of two components:
(i)~a low-complexity matrix~$\mathbf{T}$
and
(ii)~a diagonal matrix~$\mathbf{S}$.
Such matrices are given by:
\begin{align}
\label{equation-dct-approx}
\widehat{\mathbf{C}}
=
\mathbf{S} \cdot \mathbf{T}
,
\end{align}
where
\begin{align}
\label{equation-matrix-S}
\mathbf{S} = \sqrt{(\mathbf{T} \cdot {\mathbf{T}}^\top)^{-1}}
.
\end{align}
The operation $\sqrt{\cdot}$
is the matrix square root operation~\cite{Higham1987, Seber2008}.

Hereafter
\emph{low-complexity}
matrices are
referred to as $\mathbf{T}_\ast$,
where the subscript $\ast$ indicates
the considered method.
Also
\emph{DCT approximations}
are referred to as $\hat{\mathbf{C}}_\ast$.
If the subscript is absent,
then we refer to a generic low-complexity matrix or
DCT approximation.

A traditional
DCT approximation
is the signed DCT (SDCT)~\cite{Haweel2001}
which
matrix is obtained
according to:
$
\frac{1}{\sqrt{8}} \cdot \operatorname{sgn}(\mathbf{C})$,
where
$\operatorname{sgn}(\cdot)$
is the entry-wise signum function.
Therefore,
in this case,
the entries of the associated low-complexity matrix
$\mathbf{T}_\text{SDCT} = \operatorname{sgn}(\mathbf{C})$
are in $\{ 0, \pm 1 \}$.
Thus
matrix $\mathbf{T}_\text{SDCT}$
is multiplierless.

Notably,
in the past few years,
several approximations for the DCT
have been proposed
as,
for example,
the rounded DCT~(RDCT, $\mathbf{T}_\text{RDCT}$)~\cite{Cintra2010},
the modified RDCT~(MRDCT, $\mathbf{T}_\text{MRDCT}$)~\cite{Cintra2012},
the series of approximations
proposed by Bouguezel--Ahmad--Swamy~(BAS)~\cite{bas2008n, bas2008s, bas2009, bas2011, bas2013},
the Lengwehasatit--Ortega approximation~(LO, $\mathbf{T}_\text{LO}$)~\cite{ortega2004},
the approximation proposed by Pati~\emph{et~al.}~\cite{Pati2010},
and
the collection of approximations introduced in~\cite{Cintra2014}.
Most of these approximations are
orthogonal
with
low computational complexity matrix entries.
Essentially,
they are matrices
defined over the set $\{0, \pm1/2, \pm1, \pm2\}$,
with the multiplication by powers of two
implying simple bit-shifting operations.

Such approximations were demonstrated
to be competitive substitutes
for the DCT
and its related integer transformations
as
shown in~\cite{Cintra2010, Cintra2012, bas2008n, bas2008s, bas2009, bas2011, bas2013, ortega2004, Cintra2014}.
Table~\ref{tab:tfm}
illustrates some common integer transformations
linked to the DCT approximations.

\begin{table}
\centering
\footnotesize
\caption{Common 8-point low-complexity matrices associated to DCT approximations}
\label{tab:tfm}

\begin{tabular}{lcc}
\toprule
Method &
Transformation Matrix &
\\
\midrule \noalign{\smallskip}
$\mathbf{T}_\text{RDCT}$~\cite{Cintra2010} &
$\left[
\begin{smallmatrix}
1 & 1 & 1 & 1 & 1 & 1 & 1 & 1\\ 1 & 1 & 1 & 0 & 0 &-1 &-1 &-1\\ 1 & 0 & 0 &-1 &-1 & 0 & 0 & 1\\ 1 & 0 &-1 &-1 & 1 & 1 & 0 &-1\\
1 &-1 &-1 & 1 & 1 &-1 &-1 & 1\\ 1 &-1 & 0 & 1 &-1 & 0 & 1 &-1\\ 0 &-1 & 1 & 0 & 0 & 1 &-1 & 0\\ 0 &-1 & 1 &-1 & 1 &-1 & 1 & 0\\
\end{smallmatrix}
\right]$
\\\\
$\mathbf{T}_\text{BAS-2008b}$~\cite{bas2008n} &
$\left[
\begin{smallmatrix}
1 & 1 & 1 & 1 & 1 & 1 & 1 & 1 \\ 1 & 1 & 1 & 0 & 0 & -1 & -1 & -1 \\ 1 & 1 & -1 & -1 & -1 & -1 & 1 & 1 \\ 1 & 0 & -1 & 0 & 0 & 1 & 0 & -1 \\
1 & -1 & -1 & 1 & 1 & -1 & -1 & 1 \\ 1 & -1 & 1 & 0 & 0 & -1 & 1 & -1 \\ 1 & -1 & 1 & -1 & -1 & 1 & -1 & 1 \\ 1 & -1 & 1 & -1 & 1 & -1 & 1 & -1 \\
\end{smallmatrix}
\right]$
\\\\
$\mathbf{T}_\text{LO}$~\cite{ortega2004} &
$\left[
\begin{smallmatrix}
1 & 1 & 1 & 1 & 1 & 1 & 1 & 1  \\ 1 & 1 & 1 & 0 & 0 & -1 & -1 & -1 \\
1 & \frac{1}{2} & -\frac{1}{2} & -1 & -1 & -\frac{1}{2} & \frac{1}{2} & 1 \\ 1 & 0 & -1 & -1 & 1 & 1 & 0 & -1 \\
1 & -1 & -1 & 1 & 1 & -1 & -1 & 1 \\ 1 & -1 & 0 & 1 & -1 & 0 & 1 & -1 \\
\frac{1}{2} & -1 & 1 & -\frac{1}{2} & -\frac{1}{2} & 1 & -1 & \frac{1}{2} \\ 0 & -1 & 1 & -1 & 1 & -1 & 1 & 0 \\
\end{smallmatrix}
\right]$
\\\\
$\mathbf{T}_6$~\cite{Cintra2014} &
$\left[
\begin{smallmatrix}
1 & 1 & 1 & 1 & 1 & 1 & 1 & 1 \\ 2 & 1 & 1 & 0 & 0 & -1 & -1 & -2 \\ 2 & 1 & -1 & -2 & -2 & -1 & 1 & 2 \\ 1 & 0 & -2 & -1 & 1 & 2 & 0 & -1 \\
1 & -1 & -1 & 1 & 1 & -1 & -1 & 1 \\ 1 & -2 & 0 & 1 & -1 & 0 & 2 & -1 \\ 1 & -2 & 2 & -1 & -1 & 2 & -2 & 1 \\ 0 & -1 & 1 & -2 & 2 & -1 & 1 & 0 \\
\end{smallmatrix}
\right]$
\\\\
$\mathbf{T}_4$~\cite{Cintra2014} &
$\left[
\begin{smallmatrix}
1 & 1 & 1 & 1 & 1 & 1 & 1 & 1 \\ 1 & 1 & 1 & 0 & 0 & -1 & -1 & -1 \\ 1 & 1 & -1 & -1 & -1 & -1 & 1 & 1 \\ 1 & 0 & -1 & -1 & 1 & 1 & 0 & -1 \\
1 & -1 & -1 & 1 & 1 & -1 & -1 & 1 \\ 1 & -1 & 0 & 1 & -1 & 0 & 1 & -1 \\ 1 & -1 & 1 & -1 & -1 & 1 & -1 & 1 \\ 0 & -1 & 1 & -1 & 1 & -1 & 1 & 0 \\
\end{smallmatrix}
\right]$
\\
\noalign{\smallskip}
\bottomrule
\end{tabular}
\end{table}

\section{Greedy Approximations}
\label{s:method}

\subsection{Optimization Approach}

Approximate DCT matrices
are often obtained
by fully considering the exact DCT matrix~$\mathbf{C}$,
including its symmetries~\cite{Cham1989},
fast algorithms~\cite{hou87, loeffler1991},
parametrizations~\cite{fw1992},
and
numerical properties~\cite{Yuan2006}.
Usually
the
low-complexity component
of a DCT approximation
is found by solving
the following
optimization problem:
\begin{align*}
\mathbf{T}
=
\arg
\min_{\mathbf{T}'}
\operatorname{approx}
(
\mathbf{T}'
,
\mathbf{C}
)
,
\end{align*}
where
$\operatorname{approx}(\cdot,\cdot)$
is a particular approximation assessment function---such as proximity measures
and
coding performance metrics~\cite{britanak2007discrete}---and
subject
to
various constraints,
such
as orthogonality and low-complexity
of the candidate matrices~$\mathbf{T}'$.

However,
the DCT matrix
can be
understood as a stack of row vectors
$\mathbf{c}_k^\top$,
$k=1,2,\ldots,8$,
as follows:
\begin{align}
\label{equation-dct-matrix}
\mathbf{C} =
\begin{bmatrix}
\mathbf{c}_1 &
\mathbf{c}_2 &
\mathbf{c}_3 &
\mathbf{c}_4 &
\mathbf{c}_5 &
\mathbf{c}_6 &
\mathbf{c}_7 &
\mathbf{c}_8
\end{bmatrix}^\top
.
\end{align}
In the current work,
to derive an approximation for~$\mathbf{C}$,
we propose
to
individually
approximate
each of its rows
in the hope
that the set of approximate rows
generate a good approximate matrix.
Such heuristic
can be
categorized
as a greedy method~\cite{Cormen2001}.
Therefore,
our goal is to derive
a low-complexity integer matrix
\begin{align}
\label{equation-dct-approximate-matrix}
\mathbf{T} =
\begin{bmatrix}
\mathbf{t}_1 &
\mathbf{t}_2 &
\mathbf{t}_3 &
\mathbf{t}_4 &
\mathbf{t}_5 &
\mathbf{t}_6 &
\mathbf{t}_7 &
\mathbf{t}_8
\end{bmatrix}^\top
\end{align}
such that
its rows $\mathbf{t}_k^\top$,
$k=1,2,\ldots,8$,
satisfy:
\begin{align}
\label{equation-general-greedy}
\mathbf{t}_k
=
\arg
\min_{\mathbf{t} \in \mathcal{D}}
\operatorname{error}(\mathbf{t}, \mathbf{c}_k)
,
\quad
k = 1,2,\ldots,8
,
\end{align}
subject
to constraints
such
as
(i)~low-complexity of the candidate vector~$\mathbf{t}$
and
(ii)~orthogonality of the resulting matrix~$\mathbf{T}$.
The objective function
$\operatorname{error}(\cdot, \cdot)$
returns a given error measure
and
$\mathcal{D}$ is a suitable search space.

\subsection{Search Space}
\label{sec-searchspace}

In order to obtain
a low-complexity matrix~$\mathbf{T}$,
its entries
must be computationally simple~\cite{britanak2007discrete, blahut2010}.
We define the search space
as the collection of 8-point vectors
whose entries
are in a set,
say $\mathcal{P}$,
of low-complexity elements.
Therefore,
we have the search space
$\mathcal{D} = \mathcal{P}^8$.
Some choices for~$\mathcal{P}$ include:
$\mathcal{P}_1 = \{ 0, \pm1 \}$
and
$\mathcal{P}_2 = \{ 0, \pm1, \pm2 \}$.
Tables~\ref{tab:vet1} and~\ref{tab:vet2}
display
some elements
of the search spaces
$\mathcal{D}_1 = \mathcal{P}_1^8$
and
$\mathcal{D}_2 = \mathcal{P}_2^8$.
These search spaces
have
6,561 and 390,625
elements,
respectively.

\begin{table}
\centering
\footnotesize
\caption{Examples of approximated vectors for the search space~$\mathcal{D}_1$}
\label{tab:vet1}
\begin{tabular}{cc}
\toprule
$n$   &  Approximated Vector \\
\midrule \noalign{\smallskip}
1      & $\begin{bmatrix} 1 & 1 & 1 & 1 & 1 & 1 & 1 & 1 \end{bmatrix}^\top$ \\ \noalign{\smallskip}
2      & $\begin{bmatrix} 1 & 1 & 1 & 1 & 1 & 1 & 1 & -1 \end{bmatrix}^\top$ \\ \noalign{\smallskip}
3      & $\begin{bmatrix} 1 & 1 & 1 & 1 & 1 & 1 & 1 & 0 \end{bmatrix}^\top$  \\ \noalign{\smallskip}
4      & $\begin{bmatrix} 1 & 1 & 1 & 1 & 1 & 1 & -1 & 1 \end{bmatrix}^\top$ \\ \noalign{\smallskip}
\vdots & \vdots \\
6558   & $\begin{bmatrix} -1 & -1 & -1 & -1 & -1 & -1 & 1 & -1 \end{bmatrix}^\top$ \\ \noalign{\smallskip}
6559   & $\begin{bmatrix} -1 & -1 & -1 & -1 & -1 & -1 & -1 & 1 \end{bmatrix}^\top$ \\ \noalign{\smallskip}
6560   & $\begin{bmatrix} -1 & -1 & -1 & -1 & -1 & -1 & -1 & 0 \end{bmatrix}^\top$ \\ \noalign{\smallskip}
6561   & $\begin{bmatrix} -1 & -1 & -1 & -1 & -1 & -1 & -1 & -1 \end{bmatrix}^\top$ \\ \noalign{\smallskip}
\bottomrule
\end{tabular}
\end{table}

\begin{table}
\centering
\footnotesize
\caption{Examples of approximated vectors for the search space~$\mathcal{D}_2$}
\label{tab:vet2}
\begin{tabular}{cc}
\toprule
$n$   &  Approximated Vector \\
\midrule \noalign{\smallskip}
1   & $\begin{bmatrix} 2 & 2 & 2 & 2 & 2 & 2 & 2 & -1 \end{bmatrix}^\top$ \\ \noalign{\smallskip}
2   & $\begin{bmatrix} 2 & 2 & 2 & 2 & 2 & 2 & 2 & 0 \end{bmatrix}^\top$ \\ \noalign{\smallskip}
3   & $\begin{bmatrix} 2 & 2 & 2 & 2 & 2 & 2 & 2 & 1 \end{bmatrix}^\top$ \\ \noalign{\smallskip}
4   & $\begin{bmatrix} 2 & 2 & 2 & 2 & 2 & 2 & 2 & 2 \end{bmatrix}^\top$ \\ \noalign{\smallskip}
\vdots & \vdots \\
390622      &
$\begin{bmatrix} -2 & -2 & -2 & -2 & -2 & -2 & -2 & -2 \end{bmatrix}^\top$ \\ \noalign{\smallskip}
390623      &
$\begin{bmatrix} -2 & -2 & -2 & -2 & -2 & -2 & -2 & -1 \end{bmatrix}^\top$ \\ \noalign{\smallskip}
390624      &
$\begin{bmatrix} -2 & -2 & -2 & -2 & -2 & -2 & -2 & 1 \end{bmatrix}^\top$  \\ \noalign{\smallskip}
390625      &
$\begin{bmatrix} -2 & -2 & -2 & -2 & -2 & -2 & -2 & 0 \end{bmatrix}^\top$ \\ \noalign{\smallskip}
\bottomrule
\end{tabular}
\end{table}

\subsection{Objective Function}

The problem posed in~\eqref{equation-general-greedy}
requires the identification
of an error function
to quantify
the ``distance''
between the candidate row vectors
from~$\mathcal{D}$
and
the rows of the exact DCT.
Related literature
often consider
error functions
based on
matrix norms~\cite{cintra2011dct},
proximity to orthogonality~\cite{tablada2015},
and
coding performance~\cite{britanak2007discrete}.

In this work,
we propose the utilization
of
a distance
based
on the angle between vectors
as the objective function to be minimized.
Let $\mathbf{u}$
and $\mathbf{v}$
be two vectors
defined over the same Euclidean space.
The angle between vectors
is
simply given by:
\begin{align*}
\operatorname{angle}
(\mathbf{u},\mathbf{v})
=
\arccos
\left(
\frac{\langle \mathbf{u}, \mathbf{v} \rangle}{\|\mathbf{u}\| \cdot \|\mathbf{v}\|}
\right)
,
\end{align*}
where
$\langle \cdot, \cdot \rangle$
is the inner product
and
$\| \cdot \|$ indicates
the norm induced by the inner product~\cite{Strang1988}.

\subsection{Orthogonality and Row Order}

In addition,
we require that the ensemble of
rows~$\mathbf{t}_k^\top$,
$k=1,2,\ldots,8$,
must form an orthogonal set.
This is to ensure
that an orthogonal approximation
can be obtained.
As shown in~\cite{Higham1986, cintra2011},
for this property to be satisfied,
it suffices that:
\begin{align*}
\mathbf{T}
\cdot
\mathbf{T}^\top
=
[
\text{diagonal matrix}
]
.
\end{align*}
Because we aim at approximating
each of the exact DCT matrix rows individually,
the row sequential order
according to which the approximations are performed
may matter.
Notice that we approximate the rows of the DCT
based on a set of low-complexity rows, the search space.
For instance,
let us assume that we approximate the rows
in the following order:
$\bm{\wp}~=~(1,2,3,4,5,6,7,8)$.
Once we find a good approximate row for the first exact row,
i.e.,
a row vector in the search space which has the smallest angle
in relation to that exact row,
the second row is approximated considering only the row vectors in the search space
that are orthogonal to the approximation for the
first row.
After that, the third exact row is approximated
considering only the row vectors in the search space
that are orthogonal to the first and second rows already chosen.
And so on.
This procedure characterize
the greedy nature
of the proposed algorithm.

Consider now the approximation order $(4, 3, 5, 6, 1, 2, 7, 8)$,
a permutation of $\bm{\wp}$.
In this case,
we start by approximating the fourth exact row
considering the whole search space
because we are starting from it.
Hence,
the obtained approximate row might be different
from the one obtained by considering $\bm{\wp}$,
since in that case
the search space
is restricted in a different manner.

As an example,
consider the DCT matrix of length 8,
introduced in Section~2 of the manuscript.
If considering the low complexity set $\{-1, 0, 1\}$ and the approximation order (1, 2, 3, 4, 5, 6, 7, 8)
we obtain the following approximate matrix:
\begin{align*}
\left[
\begin{smallmatrix}
1 & \phantom{-}1 & \phantom{-}1 & \phantom{-}1 & \phantom{-}1 & \phantom{-}1 & \phantom{-}1 & \phantom{-}1\\
1 & \phantom{-}1 & \phantom{-}1 & \phantom{-}0 & \phantom{-}0 & -1 & -1 & -1\\
1 & \phantom{-}0 & \phantom{-}0 & -1 & -1 & \phantom{-}0 & \phantom{-}0 & \phantom{-}1\\
1 & \phantom{-}0 & -1 & -1 & \phantom{-}1 & \phantom{-}1 & \phantom{-}0 & -1\\
1 & -1 & -1 & \phantom{-}1 & \phantom{-}1 & -1 & -1 & \phantom{-}1\\
1 & -1 & \phantom{-}0 & \phantom{-}1 & -1 & \phantom{-}0 & \phantom{-}1 & -1\\
0 & -1 & \phantom{-}1 & \phantom{-}0 & \phantom{-}0 & \phantom{-}1 & -1 & \phantom{-}0\\
0 & -1 & \phantom{-}1 & -1 & \phantom{-}1 & -1 & \phantom{-}1 & \phantom{-}0\\
\end{smallmatrix}
\right].
\end{align*}
In other hand, if we consider the reverse approximation order, (8, 7, 6, 5, 4, 3, 2, 1),
we obtain the following matrix:
\begin{align*}
\left[
\begin{smallmatrix}
1 & \phantom{-}1 & \phantom{-}1 & \phantom{-}1 & \phantom{-}1 & \phantom{-}1 & \phantom{-}1 & \phantom{-}1\\
1 & \phantom{-}1 & \phantom{-}1 & \phantom{-}0 & \phantom{-}0 & -1 & -1 & -1\\
1 & \phantom{-}1 & -1 & -1 & -1 & -1 & \phantom{-}1 & \phantom{-}1\\
1 & \phantom{-}0 & -1 & -1 & \phantom{-}1 & \phantom{-}1 & \phantom{-}0 & -1\\
1 & -1 & -1 & \phantom{-}1 & \phantom{-}1 & -1 & -1 & \phantom{-}1\\
1 & -1 & \phantom{-}0 & \phantom{-}1 & -1 & \phantom{-}0 & \phantom{-}1 & -1\\
1 & -1 & \phantom{-}1 & -1 & -1 & \phantom{-}1 & -1 & \phantom{-}1\\
0 & -1 & \phantom{-}1 & -1 & \phantom{-}1 & -1 & \phantom{-}1 & \phantom{-}0\\
\end{smallmatrix}
\right].
\end{align*}
Therefore,
the row sequence considered matters for the resulting matrix.
The sequence matters
\emph{during}
the process of finding the approximate matrix.

Thus,
the row vectors $\mathbf{c}_k^\top$
from the exact matrix must be
approximated in all possibles
sequences.
For a systematic procedure,
all the $8! = 40320$
possible permutations
of the sequence~$\wp$
must be considered.
Let $\bm{\wp}_m$,
$m=1,2,\ldots,40320$,
be the resultant sequence
that determines
the $m$th permutation;
e.g.
$\bm{\wp}_{1250} = (1,     3,     7,     6,     5,     4,     8,     2)$.
The particular elements of a sequence
are denoted by
$\bm{\wp}_m(k)$, $k=1,2,\ldots,8$.
Then,
for the given example above,
we have
$\bm{\wp}_{1250}(2) = 3$.

\subsection{Proposed Optimization Problem}

Considering the above discussion,
we can re-cast~\eqref{equation-general-greedy}
in more precise terms.
For each permutation sequence $\bm{\wp}_m$,
we have the following
optimization problem:
\begin{align}
\label{equation-greedy}
\mathbf{t}_{\bm{\wp}_m(k)} =
\arg
\min_{\mathbf{d} \in \mathcal{D}}
\operatorname{angle}
(\mathbf{c}_{\bm{\wp}_m(k)}, \mathbf{d})
,
\quad
k = 1, 2, \ldots, 8,
\end{align}
subject to:
\begin{align*}
\langle
\mathbf{t}_{\bm{\wp}_m(i)}
,
\mathbf{t}_{\bm{\wp}_m(j)}
\rangle
=
0
,
\quad
i\neq j
,
\end{align*}
$m = 1,2,\ldots,40320$
and
a fixed search space $\mathcal{D} \in \{ \mathcal{D}_1, \mathcal{D}_2 \}$.
For each $m$,
the solution of the above problem
returns eight vectors,
$\mathbf{t}_{\bm{\wp}_m(1)}^\top,
\mathbf{t}_{\bm{\wp}_m(2)}^\top,
\ldots,
\mathbf{t}_{\bm{\wp}_m(8)}^\top$,
that are used
as the rows
of the desired low-complexity matrix.
Note that each sequence $\bm{\wp}_m$
may result in a different solution.
Effectively,
there are $8!=40320$ problems to be solved.
In principle,
each permutation~$\bm{\wp}_m$
can furnish
a different matrix.

Because the search space is relatively small,
we solved~\eqref{equation-greedy}
by means of exhaustive search.
Although simple,
such approach
ensures
the attainment of a solution
and
avoids
convergence issues~\cite{Cormen2001}.
Figure~\ref{alg:approx}
shows the pseudo-code for the adopted procedure
to solve~\eqref{equation-greedy}.
It is important to highlight that
although the proposed formulation
is applicable to arbitrary transform lengths,
it may not be computationally feasible.
For this reason,
we restrict our analysis
to the 8-point case.
Section~\ref{s:video}
discusses an alternative form
of generating higher order DCT approximations.

\begin{figure}

\hrulefill

 \begin{algorithmic}[1]
   \Procedure{abmApprox}{$\mathbf{C}$, $\bm{\wp}$, $\mathcal{D}$}

   \State $\textbf{approximations} \leftarrow$ null 3D matrix of size $8 \times 8 \times n$;

   \For{$m \leftarrow 1,\vert \bm{\wp} \vert$}
     \State $\bm{\wp}_m \leftarrow \bm{\wp}(m,:) $
     \For {$k \leftarrow 1, 2, \ldots, 8$}
	\State $\theta_{min} \leftarrow 2\pi$;
	\State $index \leftarrow 1$;
	\For{$i \leftarrow 1, 2, \ldots, \vert \mathcal{D} \vert$}
	  \State $aux \leftarrow \textbf{approximations}(:,:,m) \cdot (\mathcal{D}(i,:))^\top$

	    \If{$\operatorname{sum}(aux) = 0$}

		\State $\theta \leftarrow \operatorname{angle}(\mathbf{C}(\bm{\wp}_m(k),:), \mathcal{D}(i,:));$

		\If{$\theta <  \theta_{min}$}
		  \State $\theta_{min} \leftarrow \theta$;
		  \State $index \leftarrow i$;
		\EndIf
	  \EndIf
      \EndFor
     \State $\textbf{approximations}(\bm{\wp}_m(k),: m) \leftarrow \mathcal{D}(index,:)$;
     \EndFor
    \EndFor
   \EndProcedure
 \end{algorithmic}
\hrulefill

 \caption{Algorithm for the proposed method.}
 \label{alg:approx}

\end{figure}

\section{Results and Evaluation}
\label{s:measures}

In this section,
we apply the proposed method
aiming at the derivation
of
new approximations
for the 8-point DCT.
Subsequently,
we analyze and compare
the obtained matrices
with a representative set of DCT approximations
described in the literature
according to several
well-known figures of merit~\cite{salomon2007}.

\subsection{New 8-point DCT Approximations}

Considering the search spaces
$\mathcal{D}_1$
and
$\mathcal{D}_2$
(cf.
Table~\ref{tab:vet1}
and
Table~\ref{tab:vet2},
respectively),
we apply the proposed algorithm
to solve~\eqref{equation-greedy}.
Because the first and fifth rows
of the exact DCT
are trivially approximated
by the row vectors
$\begin{bmatrix} 1 & 1 & 1 & 1 & 1 & 1 & 1 & 1 \end{bmatrix}$
and $\begin{bmatrix} 1 & -1 & -1 & 1 & 1 & -1 & -1 & 1 \end{bmatrix}$,
respectively,
we limited
the search to the remaining six rows.
As a consequence,
the number of possible candidate matrices
is reduced to
$6! = 720$.
For~$\mathcal{D}_1$,
only two different matrices
were obtained,
which coincide
with
previously archived approximations,
namely:
(i)~the RDCT~\cite{Cintra2010}
and
(ii)~the matrix $\mathbf{T}_4$
introduced in~\cite{Cintra2014}.
These approximations
are depicted in Table~\ref{tab:tfm}.

On the other hand,
considering
the search space~$\mathcal{D}_2$,
the following two new matrices
were obtained:
\begin{align*}
\mathbf{T}_1 =
 \left[
\begin{smallmatrix}
1 & \phantom{-}1 & \phantom{-}1 & \phantom{-}1 & \phantom{-}1 & \phantom{-}1 & \phantom{-}1 & \phantom{-}1 \\
2 & \phantom{-}2 & \phantom{-}1 & \phantom{-}0 & \phantom{-}0 & -1 & -2 & -2  \\
2 & \phantom{-}1 & -1 & -2 & -2 & -1 & \phantom{-}1 & \phantom{-}2 \\
1 & \phantom{-}0 & -2 & -2 & \phantom{-}2 & \phantom{-}2 & \phantom{-}0 & -1  \\
1 & -1 & -1 & \phantom{-}1 & \phantom{-}1 & -1 & -1 & \phantom{-}1 \\
2 & -2 & \phantom{-}0 & \phantom{-}1 & -1 & \phantom{-}0 & \phantom{-}2 & -2  \\
1 & -2 & \phantom{-}2 & -1 & -1 & \phantom{-}2 & -2 & \phantom{-}1 \\
0 & -1 & \phantom{-}2 & -2 & \phantom{-}2 & -2 & \phantom{-}1 & \phantom{-}0  \\
\end{smallmatrix}
\right],
\end{align*}
\begin{align*}
\mathbf{T}_2 =
\left[
\begin{smallmatrix}
1 & \phantom{-}1 & \phantom{-}1 & \phantom{-}1 & \phantom{-}1 & \phantom{-}1 & \phantom{-}1 & \phantom{-}1  \\
2 & \phantom{-}1 & \phantom{-}2 & \phantom{-}0 & \phantom{-}0 & -2 & -1 & -2  \\
2 & \phantom{-}1 & -1 & -2 & -2 & -1 & \phantom{-}1 & \phantom{-}2 \\
2 & \phantom{-}0 & -2 & -1 & \phantom{-}1 & \phantom{-}2 & \phantom{-}0 & -2  \\
1 & -1 & -1 & \phantom{-}1 & \phantom{-}1 & -1 & -1 & \phantom{-}1  \\
1 & -2 & \phantom{-}0 & \phantom{-}2 & -2 & \phantom{-}0 & \phantom{-}2 & -1  \\
1 & -2 & \phantom{-}2 & -1 & -1 & \phantom{-}2 & -2 & \phantom{-}1 \\
0 & -2 & \phantom{-}1 & -2 & \phantom{-}2 & -1 & \phantom{-}2 & \phantom{-}0  \\
\end{smallmatrix}
\right].
\end{align*}

According to~\eqref{equation-dct-approx}
and~\eqref{equation-matrix-S},
the above
\emph{low-complexity}
matrices
$\mathbf{T}_1$ and $\mathbf{T}_2$
can be modified
to provide orthogonal transformations
$\widehat{\mathbf{C}}_1$
and
$\widehat{\mathbf{C}}_2$.
The selected orthogonalization procedure
is
based on the polar decomposition
as described in~\cite{Higham1986, cintra2011, Higham1988}.
Thus,
the
\emph{orthogonal DCT approximations}
associated to the matrices~$\mathbf{T}_1$ and $\mathbf{T}_2$
are
given by
\begin{align*}
\widehat{\mathbf{C}}_1 = \mathbf{S}_1 \cdot \mathbf{T}_1
\quad
\text{and}
\quad
\widehat{\mathbf{C}}_2 = \mathbf{S}_2 \cdot \mathbf{T}_2
,
\end{align*}
where
\begin{align*}
\mathbf{S}_i
=
\sqrt{
(
\mathbf{T}_i \cdot {\mathbf{T}_i}^\top
)^{-1}
}
,
\quad i = 1, 2
.
\end{align*}
Thus,
it follows that:
\begin{align*}
\mathbf{S}_1 & = \mathbf{S}_2
 =
\operatorname{diag}
\left(
\frac{1}{\sqrt{8}}, \frac{1}{\sqrt{18}}, \frac{1}{\sqrt{20}}, \frac{1}{\sqrt{18}},
\frac{1}{\sqrt{8}}, \frac{1}{\sqrt{18}}, \frac{1}{\sqrt{20}}, \frac{1}{\sqrt{18}}
\right).
\end{align*}

Other simulations were performed
considering extended sets of elements.
In particular,
following sets
were considered:
$\{ 0, \pm1, \pm4 \}$,
$\{ 0, \pm1, \pm8 \}$,
$\{ 0, \pm1, \pm2,  \pm4 \}$,
$\{ 0, \pm1, \pm2,  \pm8 \}$,
and
$\{ 0, \pm2, \pm4,  \pm8 \}$.
Generally,
the resulting matrices
did not perform as well as the ones being proposed.
Moreover,
the associate computational cost
was consistently higher.

The number of vectors
in the search space can be calculated as
$|\mathcal{D}| = |\mathcal{P}|^8$
(cf.~Section~\ref{sec-searchspace}).
Therefore,
including more elements to $\mathcal{P}$
effects a noticeable increase
in the size of the search space.
As a consequence,
the processing time
to derive all the $6!$ candidate matrices
increases accordingly.

\subsection{Approximation Measures}

Approximations measurements are computed
between an approximate matrix
$\hat{\mathbf{C}}$
(not the low-complexity matrix $\mathbf{T}$)
relative to the exact DCT.
To evaluate the performance
of the proposed approximations,
$\widehat{\mathbf{C}}_1$ and $\widehat{\mathbf{C}}_2$,
we selected
traditional figures of merit:
(i)~total error energy ($\epsilon(\cdot)$)~\cite{cintra2011dct};
(ii)~mean square error ($\operatorname{MSE}(\cdot)$)~\cite{britanak2007discrete, Wang2009mse};
(iii)~coding gain ($C_g(\cdot)$)~\cite{britanak2007discrete, Goyal2001cg, Yasuda1991cg};
and
(iv)~transform efficiency ($\eta(\cdot)$)~\cite{britanak2007discrete}.
The MSE and total error energy
are suitable measures
to quantify
the difference
between the exact DCT
and its approximations~\cite{britanak2007discrete}.
The coding gain
and transform efficiency
are appropriate tools
to quantify
compression,
redundancy removal,
and data decorrelation
capabilities~\cite{britanak2007discrete}.
Additionally,
due the angular nature
of the objective function
required by the proposed optimization problem,
we also considered
descriptive circular statistics~\cite{mardia2009, topics2001}.
Circular statistics
allows the quantification
of
approximation error
in terms of
the angle difference
between
the row vectors
of
the approximated and exact matrix.

Hereafter
we adopt the following quantities and notation:
the interpixel correlation is $\rho = 0.95$~\cite{britanak2007discrete, Goyal2001cg, Liang2001},
$\widehat{\mathbf{C}}$ is an approximation for the DCT,
and
$\widehat{\mathbf{R}}_{\mathbf{y}} = \widehat{\mathbf{C}} \cdot \mathbf{R_x} \cdot \widehat{\mathbf{C}}^\top$,
where $\mathbf{R_x}$ is the covariance matrix of $\mathbf{x}$,
whose elements are given by
$
\rho^{|i-j|}
$,
$i, j = 1, 2, \ldots, 8$.
We detail each of these measures below.

\subsubsection{Total Energy Error}

The total energy error
is a similarity measure
given by~\cite{cintra2011dct}:
\begin{align*}
\epsilon(\widehat{\mathbf{C}})
=
\pi
\cdot
\|
\mathbf{C} - \widehat{\mathbf{C}}
\|_{\text{F}}^2,
\end{align*}
where
$\| \cdot \|_{\text{F}}$
represents
the Frobenius norm~\cite{Watkins2004}.

\subsubsection{Mean Square Error}

The MSE
of a given approximation $\widehat{\mathbf{C}}$
is furnished by~\cite{britanak2007discrete, Wang2009mse}:
\begin{align*}
\operatorname{MSE}(\widehat{\mathbf{C}})
=
\frac{1}{8}
\cdot
\operatorname{tr}
\left(
(\mathbf{C} - \widehat{\mathbf{C}})
\cdot
\mathbf{R_x} \cdot (\mathbf{C} - \widehat{\mathbf{C}})^\top
\right)
.
\end{align*}
where $\operatorname{tr}(\cdot)$ represents the trace operator~\cite{britanak2007discrete}.
The total energy error
and the mean square error
are appropriate measures
for capturing
the approximation error
in a Euclidean distance sense.

\subsubsection{Coding Gain}

The coding gain
quantifies the
energy compaction capability
and
is given
by~\cite{britanak2007discrete}:
\begin{align*}
C_g(\widehat{\mathbf{C}})
=
10
\cdot
\log_{10}
\left\{
\frac
{\frac{1}{8} \sum_{i = 1}^{8} r_{i,i}}
{\left(
\prod_{i = 1}^{8}
r_{i,i}
\cdot
\|\widehat{\mathbf{c}}_i \|^2 \right)^{1/8}}
\right\}
,
\end{align*}
where
$r_{i,i}$
is the $i$th
element
of the diagonal of
$\widehat{\mathbf{R}}_{\mathbf{y}}$~\cite{britanak2007discrete}
and
$\widehat{\mathbf{c}}_i^\top$
is the $i$th row of $\widehat{\mathbf{C}}$.

However,
as pointed in~\cite{Yasuda1991cg},
the previous definition
is suitable for
orthogonal transforms only.
For
non-orthogonal transforms,
such as
SDCT~\cite{Haweel2001}
and
MRDCT~\cite{Cintra2012},
we adopt
the unified coding gain~\cite{Yasuda1991cg}.
For
$i = 1, 2, \ldots, 8$,
let
$\widehat{\mathbf{c}}_i^\top$
and
$\widehat{\mathbf{g}}_i^\top$
be $i$th row of
$\widehat{\mathbf{C}}$ and $\widehat{\mathbf{C}}^{-1}$,
respectively.
Then,
the unified coding gain
is given by:
\begin{align*}
C^*_g(\widehat{\mathbf{C}})
=
10
\cdot
\log_{10}
\left\{
\prod_{i = 1}^{8}
\frac{1}{\sqrt[8]{A_i \cdot B_i}}
\right\}
,
\end{align*}
where
$A_i
=
\operatorname{su}
\left[
\left(
\widehat{\mathbf{c}}_i \cdot \widehat{\mathbf{c}}_i^{\top}
\right)
\odot
\mathbf{R}_x
\right]
,
$
$\operatorname{su}(\cdot)$
returns the sum
of all elements
of the input matrix,
the operator
$\odot$
represents
the element-wise product,
and
$B_i
=
\|\widehat{\mathbf{g}}_i \|^2$.

\subsubsection{Transform Efficiency}

The transform efficiency
is an alternative measure to the coding gain,
being expressed according to~\cite{britanak2007discrete}:
\begin{align*}
\eta(\widehat{\mathbf{C}})
=
\frac
{\sum_{i = 1}^8 |r_{i,i}|}
{\sum_{i = 1}^8 \sum_{j = 1}^8
|
r_{i,j}
|
}
\cdot
100
,
\end{align*}
where $r_{i,j}$
is the $(i,j)$th entry of $\widehat{\mathbf{R}}_{\mathbf{y}}$,
$i,j=1,2,\ldots,8$~\cite{britanak2007discrete}.

\subsubsection{Circular Statistics}

Because the objective function in~\eqref{equation-greedy}
is
the operator angle,
its associate values
are distributed
around the unit circle.
This type of data
is suitably analyzed
by
circular statistics tools~\cite{mardia2009, topics2001, circular1991}.
Let $\mathbf{a}$ be an arbitrary 8-point vector
and
$\mathbf{q} = \begin{bmatrix} 1 & 0 & 0 & 0 & 0 & 0 & 0 & 0 \end{bmatrix}$.
The angle function is given by~\cite{mardia2009}:
\begin{align*}
 \theta = \operatorname{angle}(\mathbf{a}', \mathbf{q}), \quad k = 1, 2, \ldots, 8,
\end{align*}
where $\mathbf{a}'= \frac{\mathbf{a}}{\| \mathbf{a} \|}$
is the normalized vector of~$\mathbf{a}$.

The mean angle
(circular mean)
is given by~\cite{circular1991, topics2001}:
\begin{align*}
\bar{\theta}
=
\begin{cases}
\arctan{(S/C)},
&
\text{if $C > 0$ and $S \geq 0$},
\\
\pi/2,
&
\text{if $C = 0$ and $S > 0$},
\\
\arctan{(S/C)} + \pi,
&
\text{if $C < 0$},
\\
\arctan{(S/C)} + 2\pi,
&
\text{if $C \geq 0$ and $S < 0$},
\\
\text{undefined},
&
\text{if $C = 0$ and $S = 0$},
\end{cases}
\end{align*}
where
$
C = \sum_i \cos(\theta_i)
$
,
$
S = \sum_i \sin(\theta_i)
$,
and
$\{\theta_i\}$ is a collection of angles.
The circular variance
is given by~\cite{mardia2009}:
\begin{align*}
V
=
1 - \frac{\sqrt{C^2 + S^2}}{8}
.
\end{align*}
The minimal variance
occurs
when all observed angles are
identical.
In this case,
we have $V = 0$.
In other hand,
the maximum variance
occurs
when the observations
are uniformly distributed
around the unit circle.
Thus,
$V = 1$~\cite{topics2001}.

Considering
the rows
of the 8-point DCT matrix
and
of a given 8-point DCT approximate matrix,
the
angle function
furnishes
the following angles, respectively:
$\theta_{\mathbf{c}_k} = \operatorname{angle}(\mathbf{c}_k, \mathbf{q})$
and
$\theta_{\mathbf{t}_k} = \operatorname{angle}(\mathbf{t}_k, \mathbf{q})$,
$k = 1, 2, \ldots, 8$
(cf.~\eqref{equation-dct-matrix} and~\eqref{equation-dct-approximate-matrix}).
In this particular case,
the mean circular difference,
which measures
the mean difference
between the pairs of angles
is defined as follows:
\begin{align*}
\bar{D}
=
\frac{1}{8^2}
\cdot
\sum_{i = 1}^8
\sum_{j = 1}^8
\left( \pi - | \pi - | \theta_{\mathbf{c}_i} - \theta_{\mathbf{t}_j} | | \right).
\end{align*}
The expression above
considers the difference
between all the possible pairs of angles.
However,
we are interested
in comparing the angle
between the $i$th row of the DCT
and the corresponding row of the approximated matrix,
i.e.,
the cases where $i = j$.
Thus we have
the modified circular mean difference
according to:
\begin{align*}
\bar{D}_{mod}
=
\frac{1}{8}
\cdot
\sum_{i = 1}^8
\left( \pi - |\pi - | \theta_{\mathbf{c}_i} - \theta_{\mathbf{t}_i} | | \right).
\end{align*}

\subsection{Results and Comparisons}

Table~\ref{tab:mesures}
shows the obtained
measurements
for all approximations
derived,
according to~\eqref{equation-dct-approx},
from the low-complexity matrices
considered in this paper.
Table~\ref{tab:ec} brings a summary of the
descriptive circular statistics.
We also included
the exact DCT and the integer DCT (IDCT)~\cite{Ohm2012}
for comparison.
The considered IDCT is the 8-point approximation
adopted in the HEVC standard~\cite{Ohm2012}.
A more detailed analysis
on the performance
of the proposed approximation
in comparison with the IDCT
is provided in Section~\ref{s:video}.
The proposed DCT approximation $\mathbf{\widehat{C}}_1$
outperforms
all competing methods
in terms of MSE, coding gain, and transform efficiency.
It also performs as the second
best
for total error energy measurement.
It is unusual for an approximation
to simultaneously
excel
in measures
based on Euclidean distance
($\epsilon$ and MSE)
as well as
in coding-based measures.
The approximation by Lengwehasatit--Ortega
($\mathbf{\widehat{C}}_\text{LO}$)~\cite{ortega2004}
achieves second best results
MSE, and $\eta$.
Because of its relatively
inferior performance,
we removed the new approximation $\mathbf{\widehat{C}}_2$
from our subsequent analysis.
Nevertheless,
$\mathbf{\widehat{C}}_2$ could outperform
the approximations
$\mathbf{\widehat{C}}_\text{BAS-2008b}$~\cite{bas2008n},
$\mathbf{\widehat{C}}_\text{BAS-2009}$~\cite{bas2009},
$\mathbf{\widehat{C}}_\text{BAS-2011}$~\cite{bas2011},
$\mathbf{\widehat{C}}_\text{BAS-2013}$~\cite{bas2013},
$\mathbf{\widehat{C}}_\text{SDCT}$~\cite{Haweel2001},
$\mathbf{\widehat{C}}_\text{MRDCT}$~\cite{Cintra2012},
and $\mathbf{\widehat{C}}^{'}_{1}$~\cite{Cintra2014}
in all measures considered,
$\mathbf{\widehat{C}}_4$~\cite{Cintra2014}
and
$\mathbf{\widehat{C}}_5$~\cite{Cintra2014}
in terms of
total error energy
and
transform efficiency,
$\mathbf{\widehat{C}}_\text{RDCT}$~\cite{Cintra2010}
in terms of
total error energy,
and
$\mathbf{\widehat{C}}_\text{BAS-2008a}$~\cite{bas2008s}
in terms of
total error energy,
MSE
and
transform efficiency.
Hereafter we focus our attention
on the proposed approximation~$\mathbf{\widehat{C}}_1$.

\begin{table}
\centering
\footnotesize
\caption{Performance measures for DCT approximations derived from low-complexity matrices.
Exact DCT measures listed for reference}
\label{tab:mesures}
\begin{tabular}{lccccc}
\toprule

Method  &
$\epsilon$ &
MSE &
$C^*_g$ &
$\eta$
 \\
\midrule
DCT~\cite{ahmed1974}                         & 0       & 0      & 8.8259 & 93.9912  \\
IDCT (HEVC)~\cite{Ohm2012} & 0.0020 & $8.66 \times 10^{-6}$ & 8.8248 & 93.8236 \\
\midrule
$\widehat{\mathbf{C}}_1$ (proposed)                 	& 1.2194  & 0.0046  & 8.6337 & 90.4615  \\
$\widehat{\mathbf{C}}_2$ (proposed)                 	& 1.2194  & 0.0127  & 8.1024 & 87.2275  \\
$\widehat{\mathbf{C}}_\text{LO}$~\cite{ortega2004}  	& 0.8695  & 0.0061  & 8.3902 & 88.7023  \\
$\widehat{\mathbf{C}}_\text{SDCT}$~\cite{Haweel2001}                    	& 3.3158  & 0.0207  & 6.0261 & 82.6190  \\
$\widehat{\mathbf{C}}_\text{RDCT}$~\cite{Cintra2010}                    	& 1.7945  & 0.0098  & 8.1827 & 87.4297  \\
$\widehat{\mathbf{C}}_\text{MRDCT}$~\cite{Cintra2012}     		  	& 8.6592  & 0.0594  & 7.3326 & 80.8969  \\
$\widehat{\mathbf{C}}_\text{BAS-2008a}$~\cite{bas2008s}  		  	& 5.9294  & 0.0238  & 8.1194 & 86.8626  \\
$\widehat{\mathbf{C}}_\text{BAS-2008b}$~\cite{bas2008n}  		  	& 4.1875  & 0.0191  & 6.2684 & 83.1734  \\
$\widehat{\mathbf{C}}_\text{BAS-2009}$~\cite{bas2009}    		  	& 6.8543  & 0.0275  & 7.9126 & 85.3799  \\
$\widehat{\mathbf{C}}_\text{BAS-2011}$~\cite{bas2011}    		  	& 26.8462 & 0.0710  & 7.9118 & 85.6419  \\
$\widehat{\mathbf{C}}_\text{BAS-2013}$~\cite{bas2013}    		  	& 35.0639 & 0.1023  & 7.9461 & 85.3138  \\
$\widehat{\mathbf{C}}^{'}_1$~\cite{Cintra2014}    & 3.3158  & 0.0208  & 6.0462 & 83.0814  \\
$\widehat{\mathbf{C}}_4$~\cite{Cintra2014}                & 1.7945  & 0.0098  & 8.1834 & 87.1567  \\
$\widehat{\mathbf{C}}_5$~\cite{Cintra2014}                & 1.7945  & 0.0100  & 8.1369 & 86.5359  \\
$\widehat{\mathbf{C}}_6$~\cite{Cintra2014}                & 0.8695  & 0.0062  & 8.3437 & 88.0594  \\
\bottomrule
\end{tabular}

\end{table}

The proposed search algorithm is greedy,
i.e.,
it makes local optimal choices
hoping to find
the global optimum solution~\cite{Cormen2001}.
Therefore,
it is not guaranteed
that the obtained solutions
are globally optimal.
This is exactly what happens here.
As can be seen in Table~\ref{tab:ec},
the proposed matrix~$\mathbf{T}_1$
is not the transformation matrix
that provides
the lowest circular mean difference
among the approximations on literature.
Despite this fact,
the proposed matrix has outstanding performance
according to the considered measures.

\begin{table}[t]

\centering
\footnotesize
\caption{Descriptive circular statistics}
\label{tab:ec}
\begin{tabular}{lccc}
\toprule
Method   & $\bar{\theta}$ & $V$ & $\bar{D}_{mod}$\\
\midrule
Exact DCT~\cite{ahmed1974}	                        &	70.53	&	0.0089	&	0	        \\
IDCT (HEVC)~\cite{Ohm2012} & 70.50 & 0.0086 & 0.0022 \\
\midrule
$\mathbf{T}_1$ (proposed)          	&	71.12	&	0.0124	&	0.0711	\\
$\mathbf{T}_2$ (proposed)            &	71.12	&	0.0124	&	0.0343	\\
$\mathbf{T}_\text{LO}$~\cite{ortega2004}     	&	70.81	&	0.0102	&	0.0483	\\
$\mathbf{T}_\text{SDCT}$~\cite{Haweel2001}   	&	69.29	&	0	        &	0.1062	\\
$\mathbf{T}_\text{RDCT}$~\cite{Cintra2010}   	&	71.98	&	0.0174	&	0.0716	\\
$\mathbf{T}_\text{MRDCT}$~\cite{Cintra2012}   	&	75.58	&	0.0392	&	0.1646	\\
$\mathbf{T}_\text{BAS-2008a}$~\cite{bas2008s}    	&	72.35	&	0.0198	&	0.1036	\\
$\mathbf{T}_\text{BAS-2008b}$~\cite{bas2008n}    	&	67.29	&	0.0015	&	0.1097	\\
$\mathbf{T}_\text{BAS-2009}$~\cite{bas2009}     	&	72.10	&	0.0183	&	0.1334	\\
$\mathbf{T}_\text{BAS-2011}$~\cite{bas2011}     	&	73.54	&	0.0265	&	0.1492	\\
$\mathbf{T}_\text{BAS-2013}$~\cite{bas2013}     	&	69.29	&	0	        &	0.1062        \\
$\mathbf{T}^{'}_1$~\cite{Cintra2014}    &	73.54	&	0.0265	&	0.0901         \\
$\mathbf{T}_4$~\cite{Cintra2014}        &	70.57	&	0.0085	&	0.0781    \\
$\mathbf{T}_5$~\cite{Cintra2014}        &	72.45	&	0.0209	&	0.0730    \\
$\mathbf{T}_6$~\cite{Cintra2014}        &	71.27	&	0.0139	&	0.0497    \\
\bottomrule
\end{tabular}
\end{table}

\begin{figure}
\centering
\includegraphics{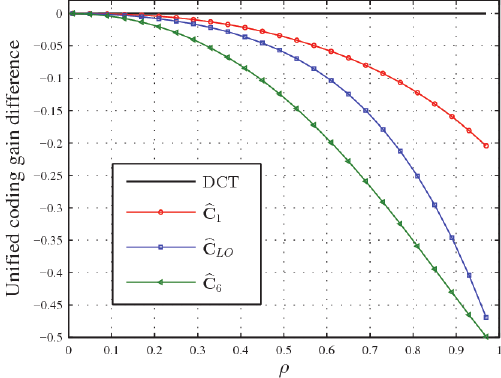}
\caption{Curves for the coding gain error of $\widehat{\mathbf{C}}_1$,
$\widehat{\mathbf{C}}_\text{LO}$,
and
$\widehat{\mathbf{C}}_6$,
relative to the exact DCT,
for $0<\rho<1$.}
\label{fig:cg_curves}
\end{figure}

Figure~\ref{fig:cg_curves}
shows the
effect of the interpixel correlation~$\rho$
on the performance of the discussed approximate transforms
as
measured by
the
unified coding gain difference
compared to the exact DCT~\cite{han2013}.
The proposed method
outperforms the competing methods
as its coding gain difference
is smaller for any choice of~$\rho$.
For highly correlated data
the coding degradation in dB
is roughly reduced by half
when the proposed approximation~$\mathbf{\widehat{C}}_1$
is considered.

\section{Fast Algorithm and Hardware Realization}
\label{s:hardware}

\subsection{Fast Algorithm}

\begin{figure*}
\centering
\includegraphics[width=0.6\linewidth]{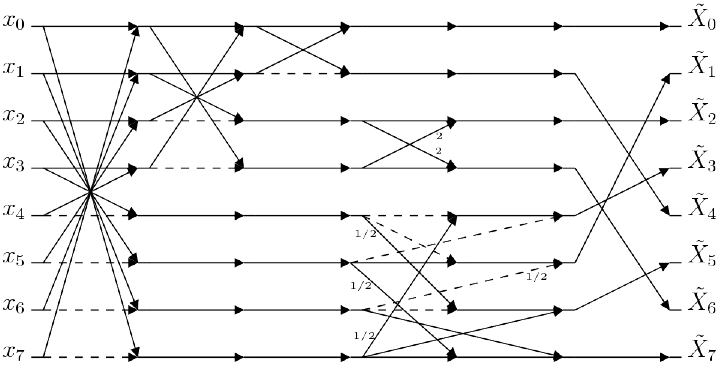}
\caption{Signal flow graph of the proposed transform,
relating the input data $x_n$, $n = 0, 1, \ldots , 7$,
to its correspondent coefficients~$\tilde{X}_k$,
$k = 0, 1, \ldots , 7$,
where $\mathbf{\tilde{X}} = \mathbf{x} \cdot \mathbf{T}_1$.
of $\mathbf{T}_1$.
Dashed arrows representing multiplication by $-1$.}
\label{fig:diag}
\end{figure*}

The direct implementation
of $\mathbf{T}_1$
requires
48~additions
and 24~bit-shifting operations.
However,
such computational cost
can be significantly reduced by means
of sparse matrix factorization~\cite{blahut2010}.
In fact,
considering
butterfly-based structures
as commonly found
in decimation-in-frequency algorithms,
such as~\cite{hou87,yip1988,rao1990},
we could derive the following factorization
for~$\mathbf{T}_1$:
\begin{align*}
\mathbf{T}_1 = \mathbf{D} \cdot \mathbf{A}_4 \cdot \mathbf{A}_3 \cdot \mathbf{A}_2 \cdot \mathbf{A}_1,
\end{align*}
where:
\begin{align*}
\mathbf{A}_1
= \left[
\begin{smallmatrix}
1 &   &   &   &   &   &   & 1\\
  & 1 &   &   &   &   & 1 &  \\
  &   & 1 &   &   & 1 &   &  \\
  &   &   & 1 & 1 &   &   &  \\
  &   &   & 1 &-1 &   &   &  \\
  &   & 1 &   &   &-1 &   &  \\
  & 1 &   &   &   &   &-1 &  \\
1 &   &   &   &   &   &   &-1\\
\end{smallmatrix}
\right]
,
\quad
\mathbf{A}_2 =
\left[
\begin{smallmatrix}
 1 &   &   & 1 &   &   &   & \\
   & 1 & 1 &   &   &   &   & \\
   & 1 &-1 &   &   &   &   & \\
 1 &   &   &-1 &   &   &   & \\
   &   &   &   & 1 &   &   & \\
   &   &   &   &   & 1 &   & \\
   &   &   &   &   &   & 1 & \\
   &   &   &   &   &   &   & 1 \\
\end{smallmatrix}
\right],
\end{align*}
\begin{align*}
\mathbf{A}_3 =
\left[
\begin{smallmatrix}
 1 & 1 &   &   &   &   &   &  \\
 1 &-1 &   &   &   &   &   &  \\
   &   & 1 &   &   &   &   &  \\
   &   &   & 1 &   &   &   &  \\
   &   &   &   & 1 &   &   &  \\
   &   &   &   &   & 1 &   &  \\
   &   &   &   &   &   & 1 &  \\
   &   &   &   &   &   &   & 1\\
\end{smallmatrix}
\right]
,
\quad
\mathbf{A}_4 =
\left[
\begin{smallmatrix}
 1 &   &   &   &   &   &   &  \\
   &   &   &   &   & \frac{1}{2} & 1  & 1\\
   &   & 1 & 2 &   &   &   &  \\
   &   &   &   &-1 &-1 &   & \frac{1}{2} \\
   & 1 &   &   &   &   &   &  \\
   &   &   &   & \frac{1}{2}  &   &-1 & 1 \\
   &   &-2 & 1 &   &   &   &  \\
   &   &   &   &-1 & 1 &-\frac{1}{2}  &  \\
\end{smallmatrix}
\right]
,
\end{align*}
and
$\mathbf{D} = \operatorname{diag}(1, 2, 1, 2, 1, 2, 1, 2)$.
Figure~\ref{fig:diag}
shows the signal flow graph~(SFG)
related to the above factorization.
The computational cost of this algorithm
is only 24~additions
and
six multiplications by two.
The
multiplications by two
are extremely simple
to be performed,
requiring only bit-shifting operations~\cite{britanak2007discrete}.
The fast algorithm proposed
requires 50\% less additions
and 75\% less bit-shifting operations
when
compared to the direct implementation.
The computational costs
of the considered methods
are shown in Table~\ref{tab:cost}.
The additive cost
of
the discussed approximations
varies from 14 to 28~additions.

In general terms,
DCT approximations
exhibit
a trade-off
between
computational cost
and
transform performance~\cite{tablada2015},
i.e.,
less complex matrices
effect poor spectral approximations~\cite{britanak2007discrete}.
Departing from this general behavior,
the proposed transformation
$\mathbf{T_1}$
has
(i)~excelling performance measures
and
(ii)~lower or similar
arithmetic cost
when compared to competing methods,
as shown in
Tables~\ref{tab:mesures}, \ref{tab:ec}, and \ref{tab:cost}.
Regarding considered performance measures,
three transformations
are consistenly placed
among the five best methods:
$\mathbf{T}_1$, $\mathbf{T}_\text{LO}$, and $\mathbf{T}_6$.
Thus,
we separate them
for hardware analysis.

\begin{table}
\centering
\footnotesize
\caption{Computational cost comparison}
\label{tab:cost}
\begin{tabular}{lccc}
\toprule
Method & Multiplications & Additions & Bit-shifts \\
\midrule
DCT~\cite{loeffler1991}             		& 11 & 29 & 0  \\
IDCT (HEVC)~\cite{Ohm2012}                             & 0 & 50 & 30 \\
\midrule
$\mathbf{T}_1$ (proposed)                     		& 0 & 24 & 6 \\
$\mathbf{T}_\text{LO}$~\cite{ortega2004}        & 0 & 24 & 2  \\
$\mathbf{T}_\text{SDCT}$~\cite{Haweel2001}          		& 0 & 24 & 0  \\
$\mathbf{T}_\text{RDCT}$~\cite{Cintra2010}           		& 0 & 22 & 0  \\
$\mathbf{T}_\text{MRDCT}$~\cite{Cintra2012}         		& 0 & 14 & 0  \\
$\mathbf{T}_\text{BAS-2008a}$~\cite{bas2008s}    			& 0 & 18 & 2  \\
$\mathbf{T}_\text{BAS-2008b}$~\cite{bas2008n}    			& 0 & 21 & 0  \\
$\mathbf{T}_\text{BAS-2009}$~\cite{bas2009}      			& 0 & 18 & 0  \\
$\mathbf{T}_\text{BAS-2011}$~\cite{bas2011}      			& 0 & 16 & 0  \\
$\mathbf{T}_\text{BAS-2013}$~\cite{bas2013}      			& 0 & 24 & 0  \\
$\mathbf{T}^{'}_1$~\cite{Cintra2014} 	& 0 & 18 & 0  \\
$\mathbf{T}_4$~\cite{Cintra2014}             	& 0 & 24 & 0  \\
$\mathbf{T}_5$~\cite{Cintra2014}             	& 0 & 24 & 4  \\
$\mathbf{T}_6$~\cite{Cintra2014}             	& 0 & 24 & 6  \\
\bottomrule
\end{tabular}
\end{table}

\subsection{FPGA Implementation}

\begin{figure*}
\centering
\subfigure[]{\includegraphics[width=0.3\linewidth]{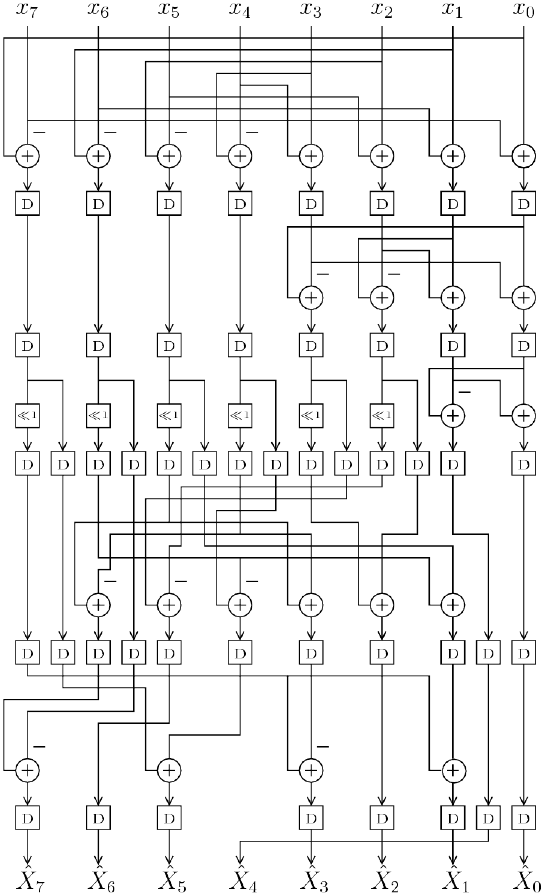}}
\quad
\subfigure[]{\includegraphics[width=0.3\linewidth]{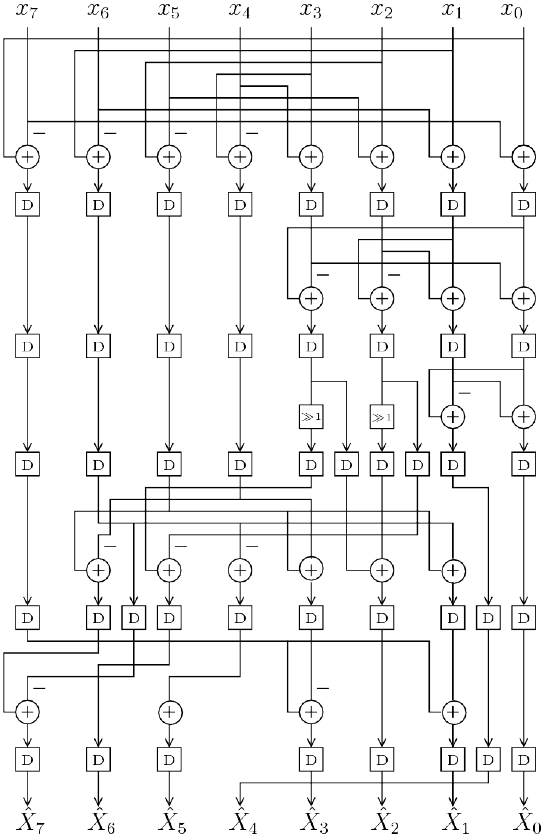}}
\quad
\subfigure[]{\includegraphics[width=0.3\linewidth]{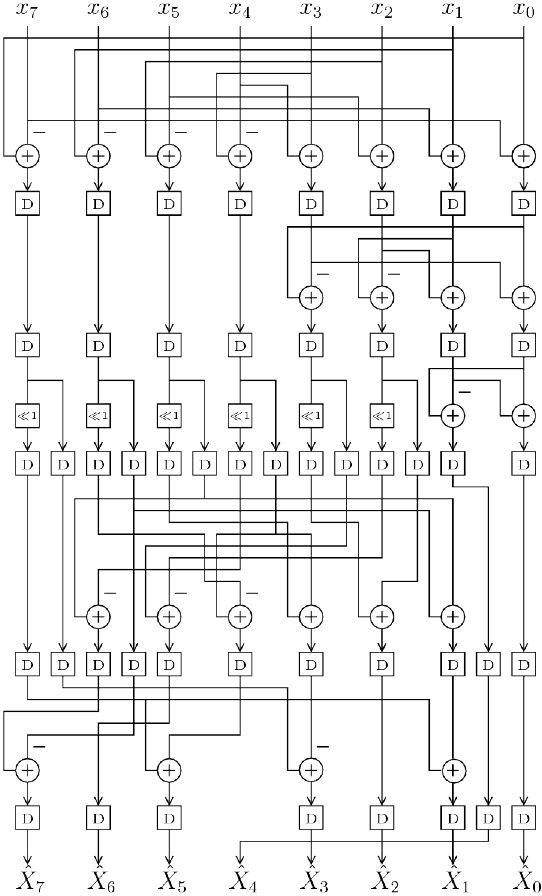}}
\caption{Architectures for
(a)~$\mathbf{T}_1$, (b)~$\mathbf{T}_\text{LO}$, and (c)~$\mathbf{T}_6$.}
\label{fig:designs}
\end{figure*}

The proposed design
along with $\mathbf{T}_\text{LO}$ and $\mathbf{T}_6$
were implemented on an FPGA chip
using the Xilinx ML605 board.
Considering hardware co-simulation
the FPGA realization was tested
with 100,000 random 8-point input test vectors.
The test vectors were generated from within the MATLAB environment
and,
using JTAG based hardware co-simulation,
routed to the physical FPGA device
where each algorithm was realized
in the reconfigurable logic fabric.
Then the computational results obtained from the FPGA algorithm
implementations were
 routed back to MATLAB memory space.
The diagrams for the designs can be seen in Figure~\ref{fig:designs}.

The metrics employed
to evaluate
the FPGA implementations
were:
configurable logic blocks (CLB),
flip-flop (FF) count,
and
critical path delay ($T_\text{cpd}$),
in ns.
The maximum operating frequency
was determined by the critical path delay
as $F_\text{max}$ = $(T_\text{cpd})^{-1}$, in $\mathrm{MHz}$.
Values were obtained from the Xilinx FPGA synthesis
and place-route tools by
accessing the \texttt{xflow.results} report file.
Using the Xilinx XPower Analyzer,
we estimated
the static ($Q_p$ in $\mathrm{W}$)
and dynamic power ($D_p$ in $\mathrm{mW/MHz}$) consumption.
In addition,
we calculated
area-time ($AT$)
and
area-time-square ($AT^2$)
figures of merit,
where
area is measured as the CLBs
and time
as the critical path delay.
The values of those metrics for each design are shown in Table~\ref{FPGAresults}.

\begin{table*}
\centering
\caption{Hardware resource consumption and power consumption using Xilinx Virtex-6 XC6VLX240T 1FFG1156 device}
\label{FPGAresults}
\begin{tabular}{lcccccccc} %
\toprule
Approximation &
CLB &
FF &
\begin{tabular}[c]{@{}c@{}}$T_\text{cpd}$ \\ ($\mathrm{ns}$)\end{tabular} &
\begin{tabular}[c]{@{}c@{}}$F_{\text{max}}$ \\ ($\mathrm{MHz}$)\end{tabular} &
\begin{tabular}[c]{@{}c@{}}$D_p$ \\ ($\mathrm{mW/GHz}$)\end{tabular} &
\begin{tabular}[c]{@{}c@{}}$Q_p$ \\ ($\mathrm{W}$)\end{tabular} &
$AT$ &
$AT^2$
\\
\midrule
{\centering $\mathbf{T}_1$ (proposed)}
&
135 & 408 & 1.750 & 571 & 2.74 & 3.471 & 236 & 413
\\
{\centering $\mathbf{T}_\text{LO}$}~\cite{ortega2004}
& 114 & 349 & 1.900 & 526 & 2.82 & 3.468 & 217 & 412
\\
{\centering $\mathbf{T}_6$}~\cite{Cintra2014}
& 125 & 389 & 2.100 & 476 & 2.57 & 3.460 & 262 & 551
\\
\bottomrule
\end{tabular}
\end{table*}

The design linked to the proposed design
approximation~$\mathbf{T}_1$
possesses the smallest $T_\text{cpd}$ among the considered methods.
Such critical path delay
allows for operations
at a $8.55\%$ and $19.96\%$ higher frequency
than the designs associated
to $\mathbf{T}_{\text{LO}}$ and $\mathbf{T}_6$,
respectively.
In terms of
area-time and are-time-square measures,
the design linked to the approximation
$\mathbf{T}_{\text{LO}}$
presents the best results,
followed by the one associated to $\mathbf{T}_1$.

\section{Computational Experiments}
\label{s:aplication}

\subsection{Still Image Compression}

\subsubsection{Experiment Setup and Results}

To evaluate
the efficiency
of the proposed transformation matrix,
we performed
a JPEG-like image compression
experiments
as described in~\cite{Cintra2014, sadhvi2014, Cintra2010}.
Input images were divided into
sub-blocks of size 8$\times$8 pixels
and submitted
to a bidimensional
(\mbox{2-D})
transformation,
such as the DCT
or one of its approximations.
Let
$\mathbf{A}$
be a sub-block
of size 8$\times$8.
The \mbox{2-D} approximate transform of~$\mathbf{A}$
is an 8$\times$8 sub-block~$\mathbf{B}$
obtained as follows~\cite{Cintra2014, cintra2011dct}:
\begin{align*}
\mathbf{B}
=
\widehat{\mathbf{C}}
\cdot
\mathbf{A}
\cdot
\widehat{\mathbf{C}}^\top
.
\end{align*}

Considering the zig-zag scan pattern
as detailed in~\cite{pao1998},
the initial $r$ elements of $\mathbf{B}$
were retained;
whereas
the remaining $(64 - r)$ elements
were discarded.
Considering 8-bit images,
this approach implies
that the fixed average bits per pixel
equals $r/8$ bits per pixel (bpp).
The previous operation results
in a matrix~$\mathbf{B}'$
populated with zeros
which is suitable for entropy encoding~\cite{Wallace1992jpeg}.
Each processed sub-block
was submitted to the corresponding~\mbox{2-D} inverse transformation
and the image was reconstructed.
For orthogonal approximations,
the~\mbox{2-D} inverse transform
is given by:
\begin{align*}
\mathbf{A} = \widehat{\mathbf{C}}^\top \cdot \mathbf{B}' \cdot \widehat{\mathbf{C}}.
\end{align*}

We considered 44 standardized images
obtained from
the `miscellaneous' volume
from USC-SIPI image bank~\cite{uscsipi},
which include
common images such as
\emph{Lena},
\emph{Boat}, \emph{Baboon},
and \emph{Peppers}.
Without loss of generality,
images were converted to 8-bit grayscale
and submitted to the
above described procedure.
The reconstructed images
were compared
with the original images
and evaluated quantitatively
according to popular figures of merit:
the mean square error (MSE)~\cite{britanak2007discrete},
the peak signal-to-noise ratio (PSNR)~\cite{salomon2007}
and
the structural similarity index~(SSIM)~\cite{Wang2004}.
We consider
the MSE and PSNR
measures
because of its good properties
and historical usage.
However,
as discussed in~\cite{Wang2009mse},
the MSE and PSNR are
not the best measures
when it comes to
predict
human perception
of image fidelity
and quality,
for which SSIM
has been shown
to be
a better measure~\cite{Wang2004, Wang2009mse}.

\begin{figure*}
\centering
\subfigure[]{\includegraphics[width=0.3\linewidth]{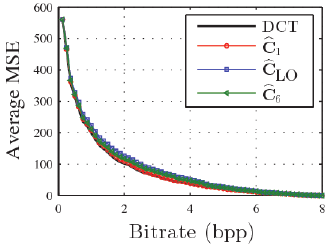}}
\subfigure[]{\includegraphics[width=0.3\linewidth]{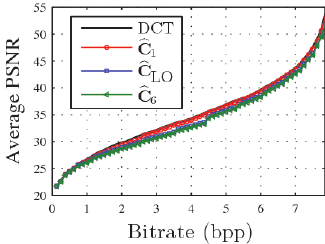}}
\subfigure[]{\includegraphics[width=0.3\linewidth]{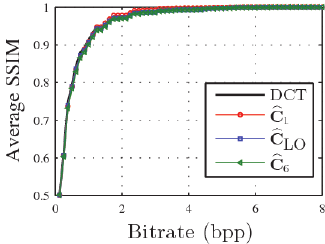}}
\caption{Curves for the average of (a)~MSE; (b)~PSNR; and (c)~SSIM corresponding to 44 images.}
\label{fig:plots1}
\end{figure*}

Figure~\ref{fig:plots1}
shows the average MSE, PSNR, and SSIM
respectively,
for the 44~images
considering~$1<r<64$ (bpp from 0 to 8) retained coefficients.
The proposed approximation $\mathbf{\widehat{C}}_1$
outperforms
$\mathbf{\widehat{C}}_\text{LO}$ and $\mathbf{\widehat{C}}_6$
in terms of MSE and PSNR
for any value of $r$.
In terms of SSIM,
$\mathbf{\widehat{C}}_1$
outperforms
$\mathbf{\widehat{C}}_6$
for any value of $r$
and
$\mathbf{\widehat{C}}_\text{LO}$
for $r\in[7,63]$.

In order to better visualize
previous curves,
we adopted
the relative difference
which is given by~\cite{higham2008functions}:
\begin{align*}
RD
=
\frac
{\mu(\mathbf{C}) - \mu(\widehat{\mathbf{C}})}
{\mu(\mathbf{C})}
,
\end{align*}
where
$\mu(\mathbf{C})$
and
$\mu(\widehat{\mathbf{C}})$
indicate
measurements
according to the exact
and approximate DCT,
respectively;
and
$\mu \in \{ \text{MSE, PSNR, SSIM}\}$.

\begin{figure*}
\centering
\subfigure[]{\includegraphics[width=0.3\linewidth]{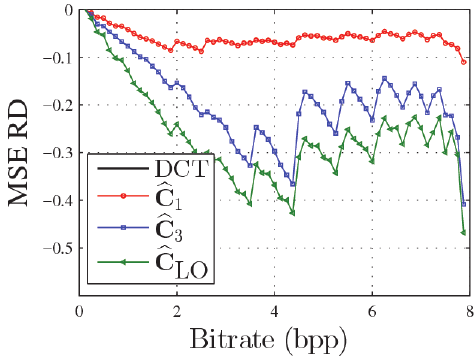}}
\subfigure[]{\includegraphics[width=0.3\linewidth]{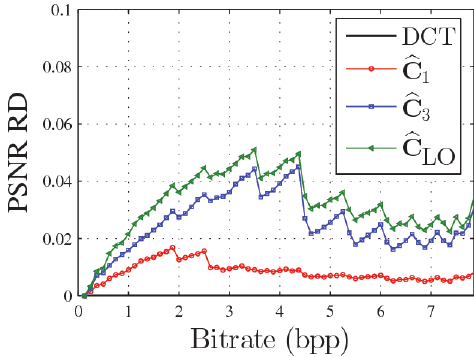}}
\subfigure[]{\includegraphics[width=0.3\linewidth]{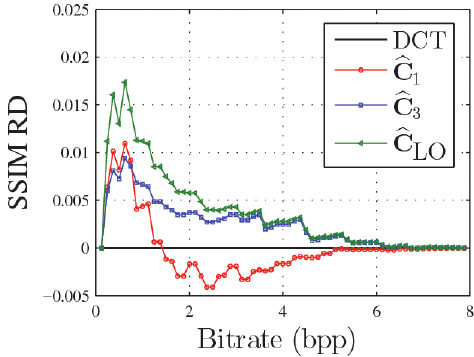}}
\caption{Relative difference curves for
(a)~MSE; (b)~PSNR; and (c)~SSIM
of $\mathbf{\widehat{C}}_1$, $\mathbf{\widehat{C}}_\text{LO}$, and $\mathbf{\widehat{C}}_6$,
relative to the exact DCT.
}
\label{fig:plots2}
\end{figure*}

The relative difference
for the
MSE, PSNR, and SSIM
are presented in Figure~\ref{fig:plots2}.
Figure~\ref{fig:plots2}(c)
shows that,
for $12<r<60$ (bpp from $1.5$ to $7.5$),
$\mathbf{\widehat{C}}_1$
outperforms not only
$\mathbf{\widehat{C}}_\text{LO}$ and $\mathbf{\widehat{C}}_6$
but the DCT itself.
To the best of our knowledge,
this particularly good behavior
was
never
described in literature,
where invariably
the performance
of DCT approximations
are routinely
bounded by the performance of the exact DCT.

\begin{figure*}
\centering
\subfigure[MSE~=~119.91, PSNR~=~27.34, SSIM~=~0.8814.]{\includegraphics[width=0.23\linewidth]{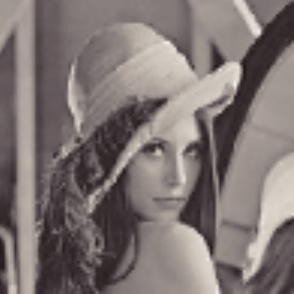}}
\subfigure[MSE~=~124.44, PSNR~=~27.18, SSIM~=~0.8767.]{\includegraphics[width=0.23\linewidth]{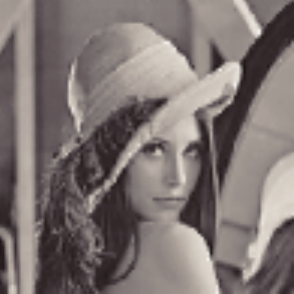}}
\subfigure[MSE~=~131.08, PSNR~=~26.95, SSIM~=~0.8781.]{\includegraphics[width=0.23\linewidth]{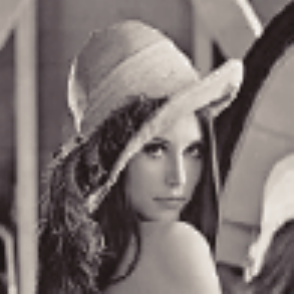}}
\subfigure[MSE~=~129.03, PSNR~=~27.02, SSIM~=~0.87.63.]{\includegraphics[width=0.23\linewidth]{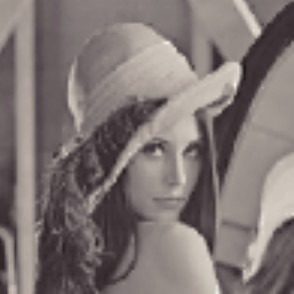}}
\caption{Compression of \texttt{Lena} using
(a)~DCT;
(b)~$\mathbf{\widehat{C}}_1$;
(c)~$\mathbf{\widehat{C}}_\text{LO}$;
and
(d)~$\mathbf{\widehat{C}}_6$ considering $r = 3$ (0.325 bpp).
}
\label{fig:lena_r3}
\end{figure*}

\begin{figure*}
\centering
\subfigure[MSE~=~27.17, PSNR~=~33.78, SSIM~=~0.9888.]{\includegraphics[width=0.23\linewidth]{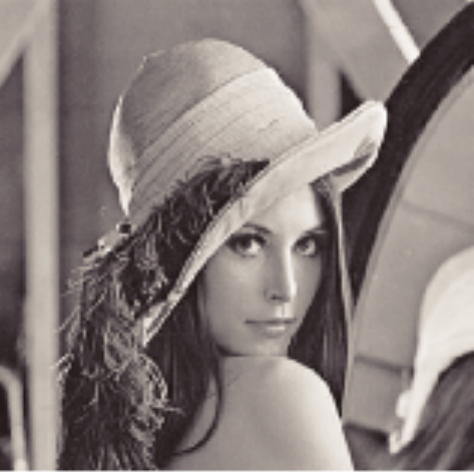}}
\subfigure[MSE~=~33.48, PSNR~=~32.88, SSIM~=~0.9893.]{\includegraphics[width=0.23\linewidth]{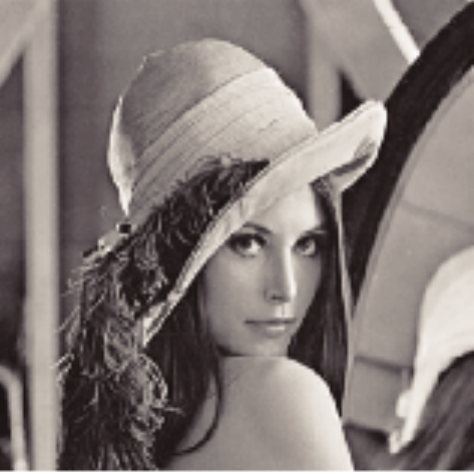}}
\subfigure[MSE~=~40.07, PSNR~=~32.10, SSIM~=~0.9849.]{\includegraphics[width=0.23\linewidth]{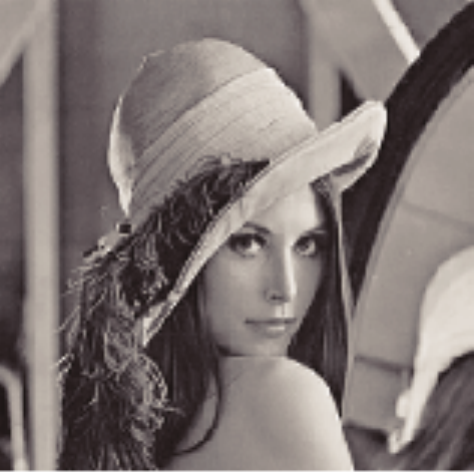}}
\subfigure[MSE~=~42.18, PSNR~=~31.87, SSIM~=~0.9844.]{\includegraphics[width=0.23\linewidth]{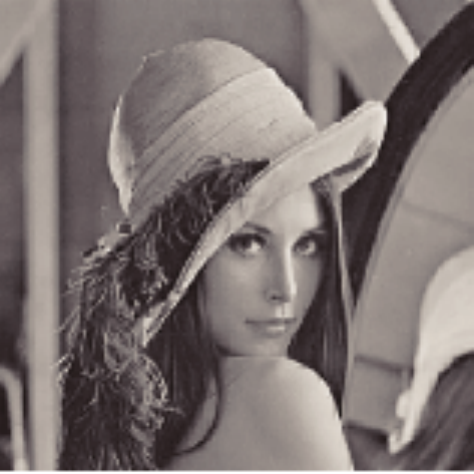}}
\caption{Compression of \texttt{Lena} using
(a)~DCT;
(b)~$\mathbf{\widehat{C}}_1$;
(c)~$\mathbf{\widehat{C}}_\text{LO}$;
and
(d)~$\mathbf{\widehat{C}}_6$ considering $r = 14$ (1.75).
}
\label{fig:lena_r14}
\end{figure*}

A qualitative evaluation
is provided in
Figures~\ref{fig:lena_r3}
and~\ref{fig:lena_r14},
where
the reconstructed \texttt{Lena} images~\cite{uscsipi}
for $r = 3$ ($0.325$ bpp) and $r = 14$ ($1.75$ bpp),
respectively,
according to
the exact DCT,
$\mathbf{\widehat{C}}_1$,
$\mathbf{\widehat{C}}_\text{LO}$,
and $\mathbf{\widehat{C}}_6$
are shown.
As expected from the results
shown in Figure~\ref{fig:plots2}(c),
for a bitrate lower than $1.5$,
the proposed approximate transform matrix
is not the one that performs the best (Figure~\ref{fig:lena_r3}),
although
the results are very similar
to the ones furnished by the exact DCT.
For a bitrate value larger than $1.5$,
Figure~\ref{fig:lena_r14} demonstrates
a situation were the proposed approximation
overcomes the other transforms,
including the DCT.
In both cases,
the visual difference
between the DCT
and the proposed aproximate transform matrix
is very small.

\subsubsection{Discussion}

The obtained approximation was capable of outperforming
the DCT under the above described conditions.
We think that this is relevant,
because it directly offers a counter-example
to the belief that
the coding performance of an approximation
is supposed to always be inferior to the DCT.
The theoretical background
that leads the optimal performance of the DCT
is based on assumption that the considered
images must follow the Markov-1 processes
with high correlation ($\rho\to 1$).
In practice,
natural images tend to fit under this assumption,
but at lower correlation values
the ideal case ($\rho\to1$) may not necessarily
be met as strongly.
For instance,
the average correlation
of the considered image set
was roughly 0.86.
This practical deviation from the ideal case ($\rho\to1$)
may also play a role
in explaining our results.

Finding matrix approximation
as
described in this work is
a
computational
and numerical task.
To the best of our knowledge,
we cannot identify any methodology
that could furnish
necessary mathematical tools
to design optimal approximations
for image compression
in an \emph{a~priori} manner,
i.e., before
search space approaches,
optimization problem solving, or numerical simulation.
In~\cite{britanak2007discrete,rao1990, tablada2015,coutinho2015, chan2006minimal, Dimitrov1998},
a large number of methods is listed;
all of them aim at good approximations.
Although
the problem of matrix approximation
is quite simple to state,
it is also
very tricky and offers several non-linearities
when combined to a more sophisticate system,
such as image compression codecs.
Finding low-complexity matrices
can be categorized as an integer optimization problem.
Thus,
navigating in the low-complexity matrix search space
might generate non-trivial performance curves,
usually leading to discontinuities,
which seems to be the case
of the proposed approximations matrix $\mathbf{T}_1$.
The navigation path through the search space
is highly dependent on
the search method and its objective function.
Although
approximation methods are very likely to provide
reasonable approximations,
it is also very hard to tell beforehand
whether a given approximation method is capable of
furnishing extremely good results
capable of outperforming the state of the art.
Only after experimentation with the obtained approximations
one may know better.
In particular,
this work
advances
an optimization setup
based on a geometrically intuitive objective function
(angle between vectors)
that could find good matrices
as demonstrated by our \emph{a~posteriori} experiments.

\subsection{Video Coding}
\label{s:video}

In order to assess the proposed transform
$\mathbf{\widehat{C}}_1$
as a tool for video coding,
we embedded it into
a public available HEVC reference software~\cite{refsoft}.
The HEVC
presents several improvements
relative to its predecessors~\cite{Sullivan2012}
and
aims at
providing
high compression rates~\cite{pourazad2012}.
Differently from other standards
(cf. Section~\ref{s:introduction}),
HEVC employs not only an 8-point integer DCT~(IDCT)
but also transforms of size
4, 16, and 32~\cite{Ohm2012}.
Such feature effects a series of optimization routines
allowing
the processing
of
big smooth or textureless areas~\cite{pourazad2012}.

The computational search used in the proposed method proved
feasible for the 8-point case.
However,
as $N$ increases,
the size of the search space
increases very quickly.
For the case $N = 16$,
if considering the low complexity set $\mathcal{P}=\{-1, 0, 1\}$,
we have a search space with $3^{16} \approx 4.3\times10^7$ elements.
For this blocklength,
The proposed algorithm
takes about 6~minutes to find
an approximation for a fixed row sequence,
considering a machine with the following specifications:
16-core 2.4 GHZ Intel(R) Xeon(R) CPU E5-2630 v3,
with 32~GB RAM running Ubuntu 16.04.3 LTS 64-bit.
Therefore,
since we have $N!$ matrices to be generated,
to run the whole algorithm
would take approximately $16! \times 6$ minutes.
In other words,
the computation time
would take an extremely large amount of time.
Thus,
for now,
we limited the scope of our computational search
to 8-point approximations.

For this reason,
aiming to derive
large blocklength transforms for
HEVC embedding,
we submitted
the proposed
transform  matrix $\mathbf{T}_1$
to
the Jridi--Alfalou--Meher~(JAM) scalable algorithm~\cite{Jridi2015}.
Such method resulted
in
16- and 32-point
versions of the proposed matrix $\mathbf{T}_1$
that are suitable for the sought video experiments.
Although the JAM algorithm is similar to
Chen's DCT~\cite{Chen1977},
it exploits redundancies allowing concise and high parallelizable hardware implementations~\cite{Jridi2015}.
From a low-complexity $N/2$-point transform,
the JAM algorithm
generates an
$N\times N$ matrix transformation
by combining two
instantiations of the smaller one.
The larger $N$-point transform is recursively defined by:
\begin{align}
\label{eq:jam}
 {\mathbf{T}}_{(N)} = \frac{1}{\sqrt{2}} \mathbf{M}^\text{per}_N \begin{bmatrix}
{\mathbf{T}}_{\left(\frac{N}{2}\right)} &  \mathbf{Z}_{\frac{N}{2}}\\
\mathbf{Z}_{\frac{N}{2}} & {\mathbf{T}}_{\left(\frac{N}{2}\right)}
\end{bmatrix} \mathbf{M}^\text{add}_N,
\end{align}
where
$\mathbf{Z}_{\frac{N}{2}}$
is a matrix of order $N/2$ with all zeroed entries.
Matrices
$\mathbf{M}^\text{add}_N$ and $\mathbf{M}^\text{per}_N$ are, respectively,
obtained according to:
\begin{align*}
\mathbf{M}^\text{add}_N =  \begin{bmatrix}
\mathbf{I}_{\frac{N}{2}} &  \mathbf{\bar{I}}_{\frac{N}{2}}\\
\mathbf{\bar{I}}_{\frac{N}{2}} & -\mathbf{I}_{\frac{N}{2}}
\end{bmatrix}
\end{align*}
and
\begin{align*}
 \mathbf{M}^\text{per}_N =  \begin{bmatrix}
\mathbf{P}_{N-1,\frac{N}{2}} &  \mathbf{Z}_{1,\frac{N}{2}}\\
\mathbf{Z}_{1,\frac{N}{2}} & \mathbf{P}_{N-1,\frac{N}{2}}
\end{bmatrix},
\end{align*}
where
$\mathbf{I}_{\frac{N}{2}}$ and $\mathbf{\bar{I}}_{\frac{N}{2}}$ are, respectively, the identity and counter-identity matrices of order $N/2$ and
$\mathbf{P}_{N-1,\frac{N}{2}}$
is an $(N-1) \times (N/2)$
matrix whose row vectors are defined by:
\begin{align*}
\mathbf{P}_{N-1,\frac{N}{2}}^{(i)} =
\begin{cases}
\mathbf{Z}_{1,\frac{N}{2}}, & \text{if $i = 1,3,5,\ldots,N-1$}\\
\mathbf{I}_{\frac{N}{2}}^{(i/2)}, & \text{if $i = 0,2,4,\ldots,N-2$}.
\end{cases}
\end{align*}
The scaling factor $1/\sqrt{2}$ of~(\ref{eq:jam})
can be merged into the image/video compression
quantization step.
Furthermore,~(\ref{equation-dct-approx}) can be applied to generate
orthogonal versions of larger transforms.
The computational cost of the resulting $N$-point transform
is given by twice
the number of bit-shifting operations
of the original $N/2$-point transform;
and twice the number of additions plus $N$ extra additions.
Following the described algorithm,
we obtained the 16- and 32-point
low complexity
transform matrices proposed.
More explicitly,
we obtained the following 16- and 32-point matrices,
respectively:
\begin{align*}
\mathbf{T}_{(16)}
=
\left[
\begin{smallmatrix}
1 & \phantom{-}1 & \phantom{-}1 & \phantom{-}1 & \phantom{-}1 & \phantom{-}1 & \phantom{-}1 & \phantom{-}1 & \phantom{-}1 & \phantom{-}1 & \phantom{-}1 & \phantom{-}1 & \phantom{-}1 & \phantom{-}1 & \phantom{-}1 & \phantom{-}1\\
1 & \phantom{-}1 & \phantom{-}1 & \phantom{-}1 & \phantom{-}1 & \phantom{-}1 & \phantom{-}1 & \phantom{-}1 & -1 & -1 & -1 & -1 & -1 & -1 & -1 & -1\\
2 & \phantom{-}2 & \phantom{-}1 & \phantom{-}0 & \phantom{-}0 & -1 & -2 & -2 & -2 & -2 & -1 & \phantom{-}0 & \phantom{-}0 & \phantom{-}1 & \phantom{-}2 & \phantom{-}2\\
2 & \phantom{-}2 & \phantom{-}1 & \phantom{-}0 & \phantom{-}0 & -1 & -2 & -2 & \phantom{-}2 & \phantom{-}2 & \phantom{-}1 & \phantom{-}0 & \phantom{-}0 & -1 & -2 & -2\\
2 & \phantom{-}1 & -1 & -2 & -2 & -1 & \phantom{-}1 & \phantom{-}2 & \phantom{-}2 & \phantom{-}1 & -1 & -2 & -2 & -1 & \phantom{-}1 & \phantom{-}2\\
2 & \phantom{-}1 & -1 & -2 & -2 & -1 & \phantom{-}1 & \phantom{-}2 & -2 & -1 & \phantom{-}1 & \phantom{-}2 & \phantom{-}2 & \phantom{-}1 & -1 & -2\\
1 & \phantom{-}0 & -2 & -2 & \phantom{-}2 & \phantom{-}2 & \phantom{-}0 & -1 & -1 & \phantom{-}0 & \phantom{-}2 & \phantom{-}2 & -2 & -2 & \phantom{-}0 & \phantom{-}1\\
1 & \phantom{-}0 & -2 & -2 & \phantom{-}2 & \phantom{-}2 & \phantom{-}0 & -1 & \phantom{-}1 & \phantom{-}0 & -2 & -2 & \phantom{-}2 & \phantom{-}2 & \phantom{-}0 & -1\\
1 & -1 & -1 & \phantom{-}1 & \phantom{-}1 & -1 & -1 & \phantom{-}1 & \phantom{-}1 & -1 & -1 & \phantom{-}1 & \phantom{-}1 & -1 & -1 & \phantom{-}1\\
1 & -1 & -1 & \phantom{-}1 & \phantom{-}1 & -1 & -1 & \phantom{-}1 & -1 & \phantom{-}1 & \phantom{-}1 & -1 & -1 & \phantom{-}1 & \phantom{-}1 & -1\\
2 & -2 & \phantom{-}0 & \phantom{-}1 & -1 & \phantom{-}0 & \phantom{-}2 & -2 & -2 & \phantom{-}2 & \phantom{-}0 & -1 & \phantom{-}1 & \phantom{-}0 & -2 & \phantom{-}2\\
2 & -2 & \phantom{-}0 & \phantom{-}1 & -1 & \phantom{-}0 & \phantom{-}2 & -2 & \phantom{-}2 & -2 & \phantom{-}0 & \phantom{-}1 & -1 & \phantom{-}0 & \phantom{-}2 & -2\\
1 & -2 & \phantom{-}2 & -1 & -1 & \phantom{-}2 & -2 & \phantom{-}1 & \phantom{-}1 & -2 & \phantom{-}2 & -1 & -1 & \phantom{-}2 & -2 & \phantom{-}1\\
1 & -2 & \phantom{-}2 & -1 & -1 & \phantom{-}2 & -2 & \phantom{-}1 & -1 & \phantom{-}2 & -2 & \phantom{-}1 & \phantom{-}1 & -2 & \phantom{-}2 & -1\\
0 & -1 & \phantom{-}2 & -2 & \phantom{-}2 & -2 & \phantom{-}1 & \phantom{-}0 & \phantom{-}0 & \phantom{-}1 & -2 & \phantom{-}2 & -2 & \phantom{-}2 & -1 & \phantom{-}0\\
0 & -1 & \phantom{-}2 & -2 & \phantom{-}2 & -2 & \phantom{-}1 & \phantom{-}0 & \phantom{-}0 & -1 & \phantom{-}2 & -2 & \phantom{-}2 & -2 & \phantom{-}1 & \phantom{-}0\\
\end{smallmatrix}
\right]
\end{align*}
and
\begin{align*}
&\mathbf{T}_{(32)}
=
\\
&
\tiny
\left[
\begin{smallmatrix}
1 & \phantom{-}1 & \phantom{-}1 & \phantom{-}1 & \phantom{-}1 & \phantom{-}1 & \phantom{-}1 & \phantom{-}1 & \phantom{-}1 & \phantom{-}1 & \phantom{-}1 & \phantom{-}1 & \phantom{-}1 & \phantom{-}1 & \phantom{-}1 & \phantom{-}1 & \phantom{-}1 & \phantom{-}1 & \phantom{-}1 & \phantom{-}1 & \phantom{-}1 & \phantom{-}1 & \phantom{-}1 & \phantom{-}1 & \phantom{-}1 & \phantom{-}1 & \phantom{-}1 & \phantom{-}1 & \phantom{-}1 & \phantom{-}1 & \phantom{-}1 & \phantom{-}1\\
1 & \phantom{-}1 & \phantom{-}1 & \phantom{-}1 & \phantom{-}1 & \phantom{-}1 & \phantom{-}1 & \phantom{-}1 & \phantom{-}1 & \phantom{-}1 & \phantom{-}1 & \phantom{-}1 & \phantom{-}1 & \phantom{-}1 & \phantom{-}1 & \phantom{-}1 & -1 & -1 & -1 & -1 & -1 & -1 & -1 & -1 & -1 & -1 & -1 & -1 & -1 & -1 & -1 & -1\\
1 & \phantom{-}1 & \phantom{-}1 & \phantom{-}1 & \phantom{-}1 & \phantom{-}1 & \phantom{-}1 & \phantom{-}1 & -1 & -1 & -1 & -1 & -1 & -1 & -1 & -1 & -1 & -1 & -1 & -1 & -1 & -1 & -1 & -1 & \phantom{-}1 & \phantom{-}1 & \phantom{-}1 & \phantom{-}1 & \phantom{-}1 & \phantom{-}1 & \phantom{-}1 & \phantom{-}1\\
1 & \phantom{-}1 & \phantom{-}1 & \phantom{-}1 & \phantom{-}1 & \phantom{-}1 & \phantom{-}1 & \phantom{-}1 & -1 & -1 & -1 & -1 & -1 & -1 & -1 & -1 & \phantom{-}1 & \phantom{-}1 & \phantom{-}1 & \phantom{-}1 & \phantom{-}1 & \phantom{-}1 & \phantom{-}1 & \phantom{-}1 & -1 & -1 & -1 & -1 & -1 & -1 & -1 & -1\\
2 & \phantom{-}2 & \phantom{-}1 & \phantom{-}0 & \phantom{-}0 & -1 & -2 & -2 & -2 & -2 & -1 & \phantom{-}0 & \phantom{-}0 & \phantom{-}1 & \phantom{-}2 & \phantom{-}2 & \phantom{-}2 & \phantom{-}2 & \phantom{-}1 & \phantom{-}0 & \phantom{-}0 & -1 & -2 & -2 & -2 & -2 & -1 & \phantom{-}0 & \phantom{-}0 & \phantom{-}1 & \phantom{-}2 & \phantom{-}2\\
2 & \phantom{-}2 & \phantom{-}1 & \phantom{-}0 & \phantom{-}0 & -1 & -2 & -2 & -2 & -2 & -1 & \phantom{-}0 & \phantom{-}0 & \phantom{-}1 & \phantom{-}2 & \phantom{-}2 & -2 & -2 & -1 & \phantom{-}0 & \phantom{-}0 & \phantom{-}1 & \phantom{-}2 & \phantom{-}2 & \phantom{-}2 & \phantom{-}2 & \phantom{-}1 & \phantom{-}0 & \phantom{-}0 & -1 & -2 & -2\\
2 & \phantom{-}2 & \phantom{-}1 & \phantom{-}0 & \phantom{-}0 & -1 & -2 & -2 & \phantom{-}2 & \phantom{-}2 & \phantom{-}1 & \phantom{-}0 & \phantom{-}0 & -1 & -2 & -2 & -2 & -2 & -1 & \phantom{-}0 & \phantom{-}0 & \phantom{-}1 & \phantom{-}2 & \phantom{-}2 & -2 & -2 & -1 & \phantom{-}0 & \phantom{-}0 & \phantom{-}1 & \phantom{-}2 & \phantom{-}2\\
2 & \phantom{-}2 & \phantom{-}1 & \phantom{-}0 & \phantom{-}0 & -1 & -2 & -2 & \phantom{-}2 & \phantom{-}2 & \phantom{-}1 & \phantom{-}0 & \phantom{-}0 & -1 & -2 & -2 & \phantom{-}2 & \phantom{-}2 & \phantom{-}1 & \phantom{-}0 & \phantom{-}0 & -1 & -2 & -2 & \phantom{-}2 & \phantom{-}2 & \phantom{-}1 & \phantom{-}0 & \phantom{-}0 & -1 & -2 & -2\\
2 & \phantom{-}1 & -1 & -2 & -2 & -1 & \phantom{-}1 & \phantom{-}2 & \phantom{-}2 & \phantom{-}1 & -1 & -2 & -2 & -1 & \phantom{-}1 & \phantom{-}2 & \phantom{-}2 & \phantom{-}1 & -1 & -2 & -2 & -1 & \phantom{-}1 & \phantom{-}2 & \phantom{-}2 & \phantom{-}1 & -1 & -2 & -2 & -1 & \phantom{-}1 & \phantom{-}2\\
2 & \phantom{-}1 & -1 & -2 & -2 & -1 & \phantom{-}1 & \phantom{-}2 & \phantom{-}2 & \phantom{-}1 & -1 & -2 & -2 & -1 & \phantom{-}1 & \phantom{-}2 & -2 & -1 & \phantom{-}1 & \phantom{-}2 & \phantom{-}2 & \phantom{-}1 & -1 & -2 & -2 & -1 & \phantom{-}1 & \phantom{-}2 & \phantom{-}2 & \phantom{-}1 & -1 & -2\\
2 & \phantom{-}1 & -1 & -2 & -2 & -1 & \phantom{-}1 & \phantom{-}2 & -2 & -1 & \phantom{-}1 & \phantom{-}2 & \phantom{-}2 & \phantom{-}1 & -1 & -2 & -2 & -1 & \phantom{-}1 & \phantom{-}2 & \phantom{-}2 & \phantom{-}1 & -1 & -2 & \phantom{-}2 & \phantom{-}1 & -1 & -2 & -2 & -1 & \phantom{-}1 & \phantom{-}2\\
2 & \phantom{-}1 & -1 & -2 & -2 & -1 & \phantom{-}1 & \phantom{-}2 & -2 & -1 & \phantom{-}1 & \phantom{-}2 & \phantom{-}2 & \phantom{-}1 & -1 & -2 & \phantom{-}2 & \phantom{-}1 & -1 & -2 & -2 & -1 & \phantom{-}1 & \phantom{-}2 & -2 & -1 & \phantom{-}1 & \phantom{-}2 & \phantom{-}2 & \phantom{-}1 & -1 & -2\\
1 & \phantom{-}0 & -2 & -2 & \phantom{-}2 & \phantom{-}2 & \phantom{-}0 & -1 & -1 & \phantom{-}0 & \phantom{-}2 & \phantom{-}2 & -2 & -2 & \phantom{-}0 & \phantom{-}1 & \phantom{-}1 & \phantom{-}0 & -2 & -2 & \phantom{-}2 & \phantom{-}2 & \phantom{-}0 & -1 & -1 & \phantom{-}0 & \phantom{-}2 & \phantom{-}2 & -2 & -2 & \phantom{-}0 & \phantom{-}1\\
1 & \phantom{-}0 & -2 & -2 & \phantom{-}2 & \phantom{-}2 & \phantom{-}0 & -1 & -1 & \phantom{-}0 & \phantom{-}2 & \phantom{-}2 & -2 & -2 & \phantom{-}0 & \phantom{-}1 & -1 & \phantom{-}0 & \phantom{-}2 & \phantom{-}2 & -2 & -2 & \phantom{-}0 & \phantom{-}1 & \phantom{-}1 & \phantom{-}0 & -2 & -2 & \phantom{-}2 & \phantom{-}2 & \phantom{-}0 & -1\\
1 & \phantom{-}0 & -2 & -2 & \phantom{-}2 & \phantom{-}2 & \phantom{-}0 & -1 & \phantom{-}1 & \phantom{-}0 & -2 & -2 & \phantom{-}2 & \phantom{-}2 & \phantom{-}0 & -1 & -1 & \phantom{-}0 & \phantom{-}2 & \phantom{-}2 & -2 & -2 & \phantom{-}0 & \phantom{-}1 & -1 & \phantom{-}0 & \phantom{-}2 & \phantom{-}2 & -2 & -2 & \phantom{-}0 & \phantom{-}1\\
1 & \phantom{-}0 & -2 & -2 & \phantom{-}2 & \phantom{-}2 & \phantom{-}0 & -1 & \phantom{-}1 & \phantom{-}0 & -2 & -2 & \phantom{-}2 & \phantom{-}2 & \phantom{-}0 & -1 & \phantom{-}1 & \phantom{-}0 & -2 & -2 & \phantom{-}2 & \phantom{-}2 & \phantom{-}0 & -1 & \phantom{-}1 & \phantom{-}0 & -2 & -2 & \phantom{-}2 & \phantom{-}2 & \phantom{-}0 & -1\\
1 & -1 & -1 & \phantom{-}1 & \phantom{-}1 & -1 & -1 & \phantom{-}1 & \phantom{-}1 & -1 & -1 & \phantom{-}1 & \phantom{-}1 & -1 & -1 & \phantom{-}1 & \phantom{-}1 & -1 & -1 & \phantom{-}1 & \phantom{-}1 & -1 & -1 & \phantom{-}1 & \phantom{-}1 & -1 & -1 & \phantom{-}1 & \phantom{-}1 & -1 & -1 & \phantom{-}1\\
1 & -1 & -1 & \phantom{-}1 & \phantom{-}1 & -1 & -1 & \phantom{-}1 & \phantom{-}1 & -1 & -1 & \phantom{-}1 & \phantom{-}1 & -1 & -1 & \phantom{-}1 & -1 & \phantom{-}1 & \phantom{-}1 & -1 & -1 & \phantom{-}1 & \phantom{-}1 & -1 & -1 & \phantom{-}1 & \phantom{-}1 & -1 & -1 & \phantom{-}1 & \phantom{-}1 & -1\\
1 & -1 & -1 & \phantom{-}1 & \phantom{-}1 & -1 & -1 & \phantom{-}1 & -1 & \phantom{-}1 & \phantom{-}1 & -1 & -1 & \phantom{-}1 & \phantom{-}1 & -1 & -1 & \phantom{-}1 & \phantom{-}1 & -1 & -1 & \phantom{-}1 & \phantom{-}1 & -1 & \phantom{-}1 & -1 & -1 & \phantom{-}1 & \phantom{-}1 & -1 & -1 & \phantom{-}1\\
1 & -1 & -1 & \phantom{-}1 & \phantom{-}1 & -1 & -1 & \phantom{-}1 & -1 & \phantom{-}1 & \phantom{-}1 & -1 & -1 & \phantom{-}1 & \phantom{-}1 & -1 & \phantom{-}1 & -1 & -1 & \phantom{-}1 & \phantom{-}1 & -1 & -1 & \phantom{-}1 & -1 & \phantom{-}1 & \phantom{-}1 & -1 & -1 & \phantom{-}1 & \phantom{-}1 & -1\\
2 & -2 & \phantom{-}0 & \phantom{-}1 & -1 & \phantom{-}0 & \phantom{-}2 & -2 & -2 & \phantom{-}2 & \phantom{-}0 & -1 & \phantom{-}1 & \phantom{-}0 & -2 & \phantom{-}2 & \phantom{-}2 & -2 & \phantom{-}0 & \phantom{-}1 & -1 & \phantom{-}0 & \phantom{-}2 & -2 & -2 & \phantom{-}2 & \phantom{-}0 & -1 & \phantom{-}1 & \phantom{-}0 & -2 & \phantom{-}2\\
2 & -2 & \phantom{-}0 & \phantom{-}1 & -1 & \phantom{-}0 & \phantom{-}2 & -2 & -2 & \phantom{-}2 & \phantom{-}0 & -1 & \phantom{-}1 & \phantom{-}0 & -2 & \phantom{-}2 & -2 & \phantom{-}2 & \phantom{-}0 & -1 & \phantom{-}1 & \phantom{-}0 & -2 & \phantom{-}2 & \phantom{-}2 & -2 & \phantom{-}0 & \phantom{-}1 & -1 & \phantom{-}0 & \phantom{-}2 & -2\\
2 & -2 & \phantom{-}0 & \phantom{-}1 & -1 & \phantom{-}0 & \phantom{-}2 & -2 & \phantom{-}2 & -2 & \phantom{-}0 & \phantom{-}1 & -1 & \phantom{-}0 & \phantom{-}2 & -2 & -2 & \phantom{-}2 & \phantom{-}0 & -1 & \phantom{-}1 & \phantom{-}0 & -2 & \phantom{-}2 & -2 & \phantom{-}2 & \phantom{-}0 & -1 & \phantom{-}1 & \phantom{-}0 & -2 & \phantom{-}2\\
2 & -2 & \phantom{-}0 & \phantom{-}1 & -1 & \phantom{-}0 & \phantom{-}2 & -2 & \phantom{-}2 & -2 & \phantom{-}0 & \phantom{-}1 & -1 & \phantom{-}0 & \phantom{-}2 & -2 & \phantom{-}2 & -2 & \phantom{-}0 & \phantom{-}1 & -1 & \phantom{-}0 & \phantom{-}2 & -2 & \phantom{-}2 & -2 & \phantom{-}0 & \phantom{-}1 & -1 & \phantom{-}0 & \phantom{-}2 & -2\\
1 & -2 & \phantom{-}2 & -1 & -1 & \phantom{-}2 & -2 & \phantom{-}1 & \phantom{-}1 & -2 & \phantom{-}2 & -1 & -1 & \phantom{-}2 & -2 & \phantom{-}1 & \phantom{-}1 & -2 & \phantom{-}2 & -1 & -1 & \phantom{-}2 & -2 & \phantom{-}1 & \phantom{-}1 & -2 & \phantom{-}2 & -1 & -1 & \phantom{-}2 & -2 & \phantom{-}1\\
1 & -2 & \phantom{-}2 & -1 & -1 & \phantom{-}2 & -2 & \phantom{-}1 & \phantom{-}1 & -2 & \phantom{-}2 & -1 & -1 & \phantom{-}2 & -2 & \phantom{-}1 & -1 & \phantom{-}2 & -2 & \phantom{-}1 & \phantom{-}1 & -2 & \phantom{-}2 & -1 & -1 & \phantom{-}2 & -2 & \phantom{-}1 & \phantom{-}1 & -2 & \phantom{-}2 & -1\\
1 & -2 & \phantom{-}2 & -1 & -1 & \phantom{-}2 & -2 & \phantom{-}1 & -1 & \phantom{-}2 & -2 & \phantom{-}1 & \phantom{-}1 & -2 & \phantom{-}2 & -1 & -1 & \phantom{-}2 & -2 & \phantom{-}1 & \phantom{-}1 & -2 & \phantom{-}2 & -1 & \phantom{-}1 & -2 & \phantom{-}2 & -1 & -1 & \phantom{-}2 & -2 & \phantom{-}1\\
1 & -2 & \phantom{-}2 & -1 & -1 & \phantom{-}2 & -2 & \phantom{-}1 & -1 & \phantom{-}2 & -2 & \phantom{-}1 & \phantom{-}1 & -2 & \phantom{-}2 & -1 & \phantom{-}1 & -2 & \phantom{-}2 & -1 & -1 & \phantom{-}2 & -2 & \phantom{-}1 & -1 & \phantom{-}2 & -2 & \phantom{-}1 & \phantom{-}1 & -2 & \phantom{-}2 & -1\\
0 & -1 & \phantom{-}2 & -2 & \phantom{-}2 & -2 & \phantom{-}1 & \phantom{-}0 & \phantom{-}0 & \phantom{-}1 & -2 & \phantom{-}2 & -2 & \phantom{-}2 & -1 & \phantom{-}0 & \phantom{-}0 & -1 & \phantom{-}2 & -2 & \phantom{-}2 & -2 & \phantom{-}1 & \phantom{-}0 & \phantom{-}0 & \phantom{-}1 & -2 & \phantom{-}2 & -2 & \phantom{-}2 & -1 & \phantom{-}0\\
0 & -1 & \phantom{-}2 & -2 & \phantom{-}2 & -2 & \phantom{-}1 & \phantom{-}0 & \phantom{-}0 & \phantom{-}1 & -2 & \phantom{-}2 & -2 & \phantom{-}2 & -1 & \phantom{-}0 & \phantom{-}0 & \phantom{-}1 & -2 & \phantom{-}2 & -2 & \phantom{-}2 & -1 & \phantom{-}0 & \phantom{-}0 & -1 & \phantom{-}2 & -2 & \phantom{-}2 & -2 & \phantom{-}1 & \phantom{-}0\\
0 & -1 & \phantom{-}2 & -2 & \phantom{-}2 & -2 & \phantom{-}1 & \phantom{-}0 & \phantom{-}0 & -1 & \phantom{-}2 & -2 & \phantom{-}2 & -2 & \phantom{-}1 & \phantom{-}0 & \phantom{-}0 & \phantom{-}1 & -2 & \phantom{-}2 & -2 & \phantom{-}2 & -1 & \phantom{-}0 & \phantom{-}0 & \phantom{-}1 & -2 & \phantom{-}2 & -2 & \phantom{-}2 & -1 & \phantom{-}0\\
0 & -1 & \phantom{-}2 & -2 & \phantom{-}2 & -2 & \phantom{-}1 & \phantom{-}0 & \phantom{-}0 & -1 & \phantom{-}2 & -2 & \phantom{-}2 & -2 & \phantom{-}1 & \phantom{-}0 & \phantom{-}0 & -1 & \phantom{-}2 & -2 & \phantom{-}2 & -2 & \phantom{-}1 & \phantom{-}0 & \phantom{-}0 & -1 & \phantom{-}2 & -2 & \phantom{-}2 & -2 & \phantom{-}1 & \phantom{-}0\\
\end{smallmatrix}
\right].
\end{align*}
The resulting approximations for the above low-complexity
matrices
can be found from~\eqref{equation-dct-approx}
and~\eqref{equation-matrix-S}.
The diagonal matrices implied by~\eqref{equation-matrix-S}
are
$
\mathbf{D}_{(16)}
=
4\cdot
\left[
\begin{smallmatrix}
1 & 0 \\ 0 & 1
\end{smallmatrix}
\right]
\otimes
\operatorname{diag}(4, 9, 10, 9)
\otimes
\left[
\begin{smallmatrix}
1 & 0 \\ 0 & 1
\end{smallmatrix}
\right]
$
and
$
\mathbf{D}_{(32)}
=
2
\cdot
\mathbf{D}_{(16)}
\otimes
\left[
\begin{smallmatrix}
1 & 0 \\ 0 & 1
\end{smallmatrix}
\right]
$,
respectively,
where
$\otimes$ is the Kronecker product~\cite{Seber2008}.

\begin{figure}[ht]
\centering
\includegraphics[scale=.6]{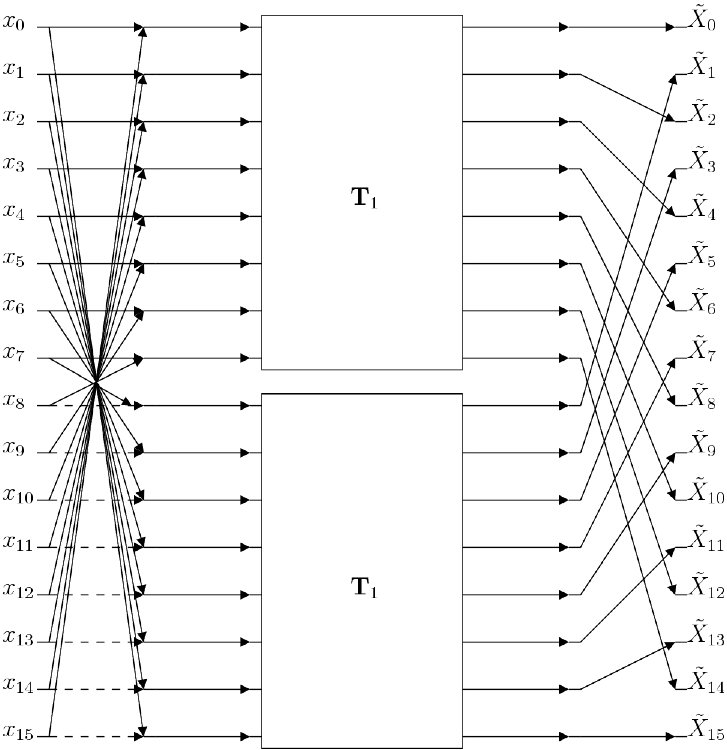}
\caption{
SFG for the proposed 16-point low complexity
transform matrix.}
\label{fig:approx16sfg}
\end{figure}

\begin{figure}[ht]
\centering
\includegraphics[scale=.6]{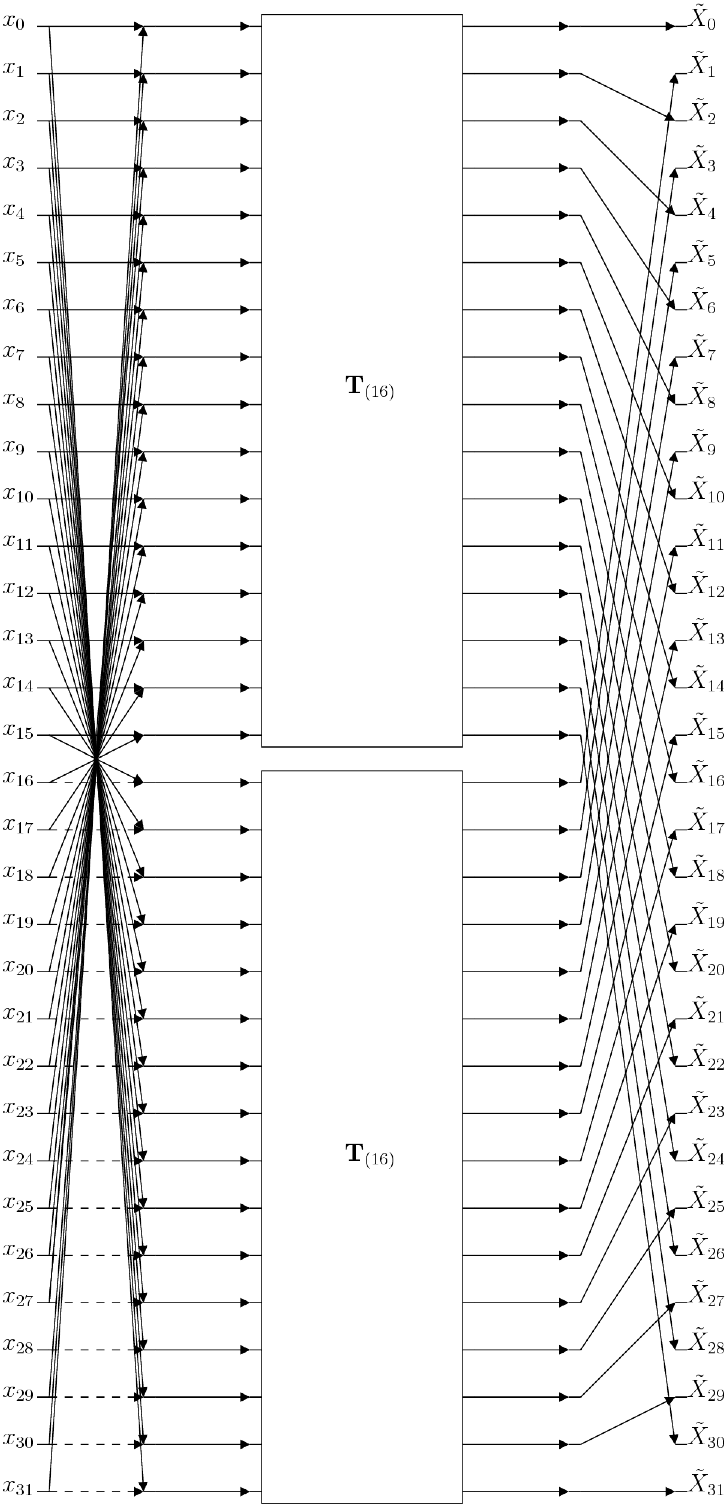}
\caption{
SFG for the proposed 32-point low complexity
transform matrix,
where $\mathbf{T}_{(16)}$
is the 16-point matrix
presented in Figure~\ref{fig:approx16sfg}.}
\label{fig:approx32sfg}
\end{figure}

Figures~\ref{fig:approx16sfg}
and \ref{fig:approx32sfg}
display the SFG
for the low-complexity transform matrices
$\mathbf{T}_{(16)}$ and $\mathbf{T}_{(32)}$
derived from $\mathbf{T}_1$.

Table~\ref{tab:hevccost} lists
the computational costs of the proposed transform for sizes $N = 8, 16, 32$
compared to an efficient implementation of the IDCT~\cite{meher2014}.

\begin{table}[]
\centering
\caption{Computational cost comparison for 8-, 16-, and 32-point transforms embedded in HEVC reference software}
\label{tab:hevccost}
\begin{tabular}{ccccc}
\toprule
\multirow{2}{*}{$N$} & \multicolumn{2}{c}{IDCT~\cite{meher2014}} & \multicolumn{2}{c}{Proposed transform} \\
\cmidrule{2-5}
& Additions  & Bit-shifts
& Additions  & Bit-shifts
\\
\midrule
8   & 50   & 30   & 24   & 6  \\
16  & 186  & 86   & 64   & 12 \\
32  & 682  & 278  & 160  & 24 \\
\bottomrule
\end{tabular}
\end{table}

In our experiments,
the
original 8-, 16-, and 32-point
integer transforms of HEVC
were substituted
by
$\mathbf{\widehat{C}}_1$
and its scaled versions.
The original 4-point transform
was kept unchanged because it is already a very
low-complexity transformation.
We encoded the first 100 frames of one video sequence of each A to F class in accordance with the common test conditions (CTC) documentation~\cite{ctconditions2013}. Namely we used the 8-bit videos:
\texttt{PeopleOnStreet} (2560$\times$1600 at 30~fps),
\texttt{BasketballDrive} (1920$\times$1080 at 50~fps),
\texttt{RaceHorses} (832$\times$480 at 30~fps),
\texttt{BlowingBubbles} (416$\times$240  at 50~fps),
\texttt{KristenAndSara} (1280$\times$720  at 60~fps),
and
\texttt{BasketbalDrillText} (832$\times$480  at 50~fps).
As suggested in~\cite{Jridi2015}, all the test parameters were set according to the CTC documentation.
We tested the proposed transforms in
All Intra (AI),
\texttt{Random Access} (RA),
\texttt{Low Delay B} (LD-B),
and
\texttt{Low Delay P} (LD-P)
configurations,
all in the \texttt{Main} profile.

\begin{figure*}
\centering
\subfigure[]{\includegraphics[scale=.5]{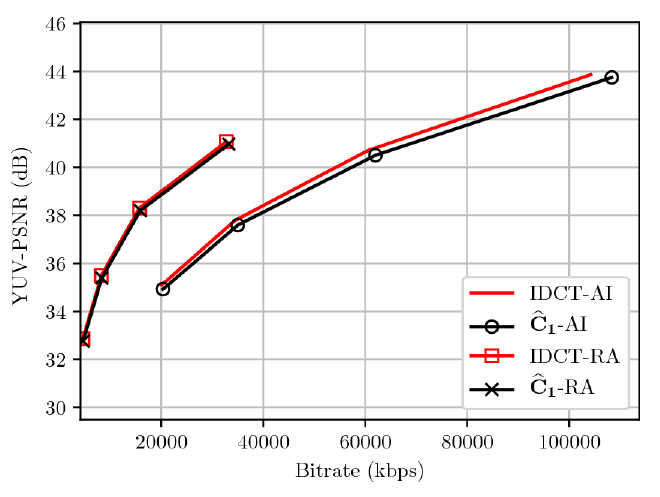}}
\subfigure[]{\includegraphics[scale=.5]{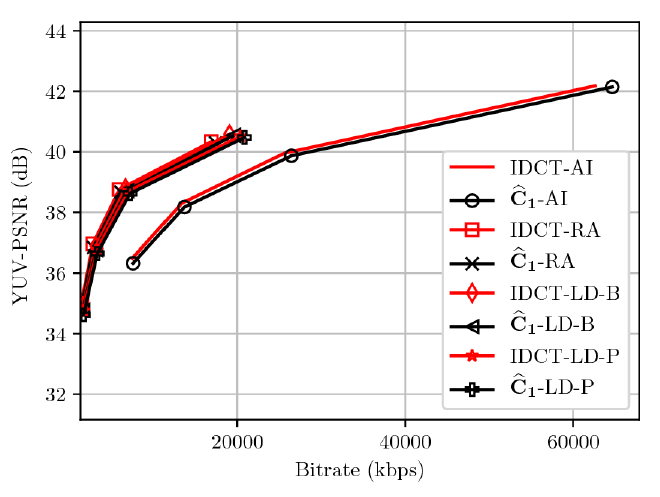}}
\subfigure[]{\includegraphics[scale=.5]{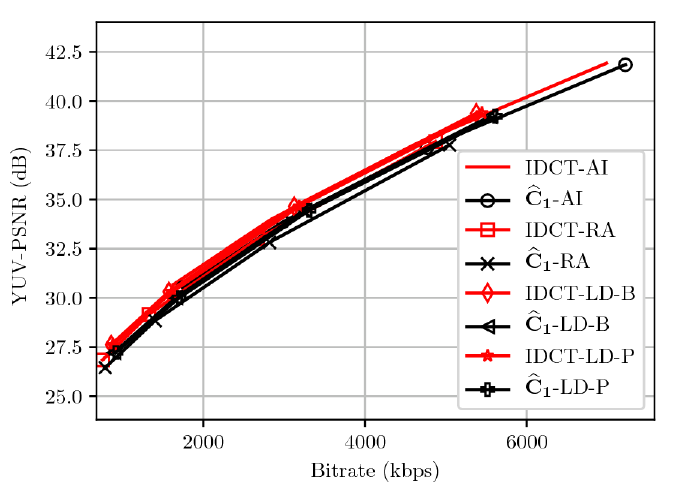}}
\subfigure[]{\includegraphics[scale=.5]{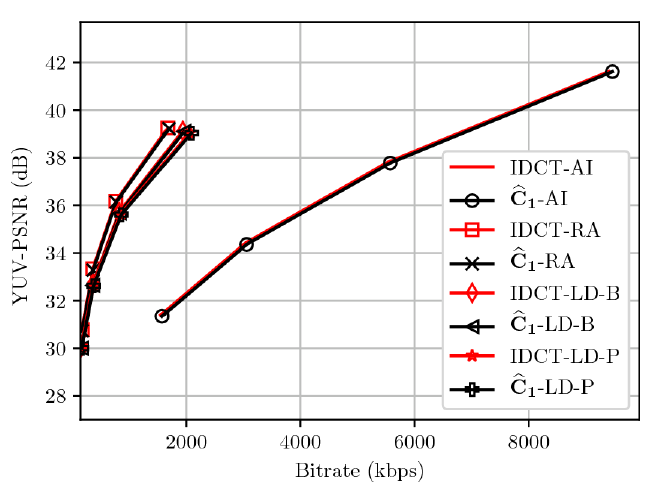}}
\subfigure[]{\includegraphics[scale=.5]{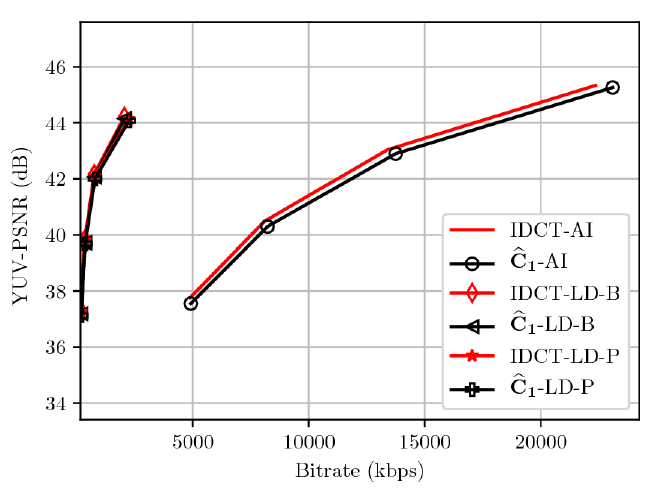}}
\subfigure[]{\includegraphics[scale=.5]{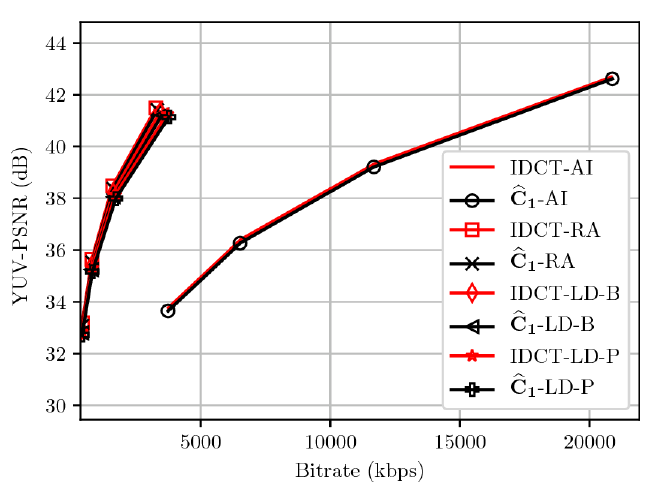}}
\caption{%
Rate distortion curves of the modified HEVC software for test sequences:
(a)~\texttt{PeopleOnStreet},
(b)~\texttt{BasketballDrive},
(c)~\texttt{RaceHorses},
(d)~\texttt{BlowingBubbles},
(e)~\texttt{KristenAndSara},
and
(f)~\texttt{BasketbalDrillText}.
}
\label{fig:allbitrates}
\end{figure*}

We selected the frame-by-frame MSE and PSNR~\cite{Ohm2012} for each YUV color channel as figures of merit.
Then, for all test videos,
we computed the rate distortion (RD) curve considering
the recommended quantization parameter (QP) values, i.e. 22, 27, 32, and 37~\cite{ctconditions2013}.
The resulting RD curves are depicted in Figure~\ref{fig:allbitrates}.
We have also measured the Bj{\o}ntegaard's delta PSNR (BD-PSNR)
and delta rate (BD-Rate)~\cite{bjontegaard2001, Hanhart2014}
for the modified HEVC software.
These values are summarized in Table~\ref{tab:bdpsnrbdrate}.
We demonstrate that replacing the IDCT by the proposed transform
and its scaled versions
results in a loss in quality of at most 0.47dB for the AI configuration, which corresponds to an increase of 5.82\% in bitrate.
Worst performance for the other configurations---RA, LD-B, and LD-P---are found
for the \texttt{KristenAndSara} video sequence,
where approximately 0.55dB are lost if compared to the original HEVC implementation.

\begin{figure*}
\centering
\subfigure[$\mbox{MSE-Y} = 10.4097$,
$\mbox{MSE-U} = 3.5872$,
$\mbox{MSE-V} = 3.3079$,
$\mbox{PSNR-Y} = 37.9564$,
$\mbox{PSNR-U} = 42.5832$,
$\mbox{PSNR-V} = 42.9353$]{\includegraphics[scale=.15]{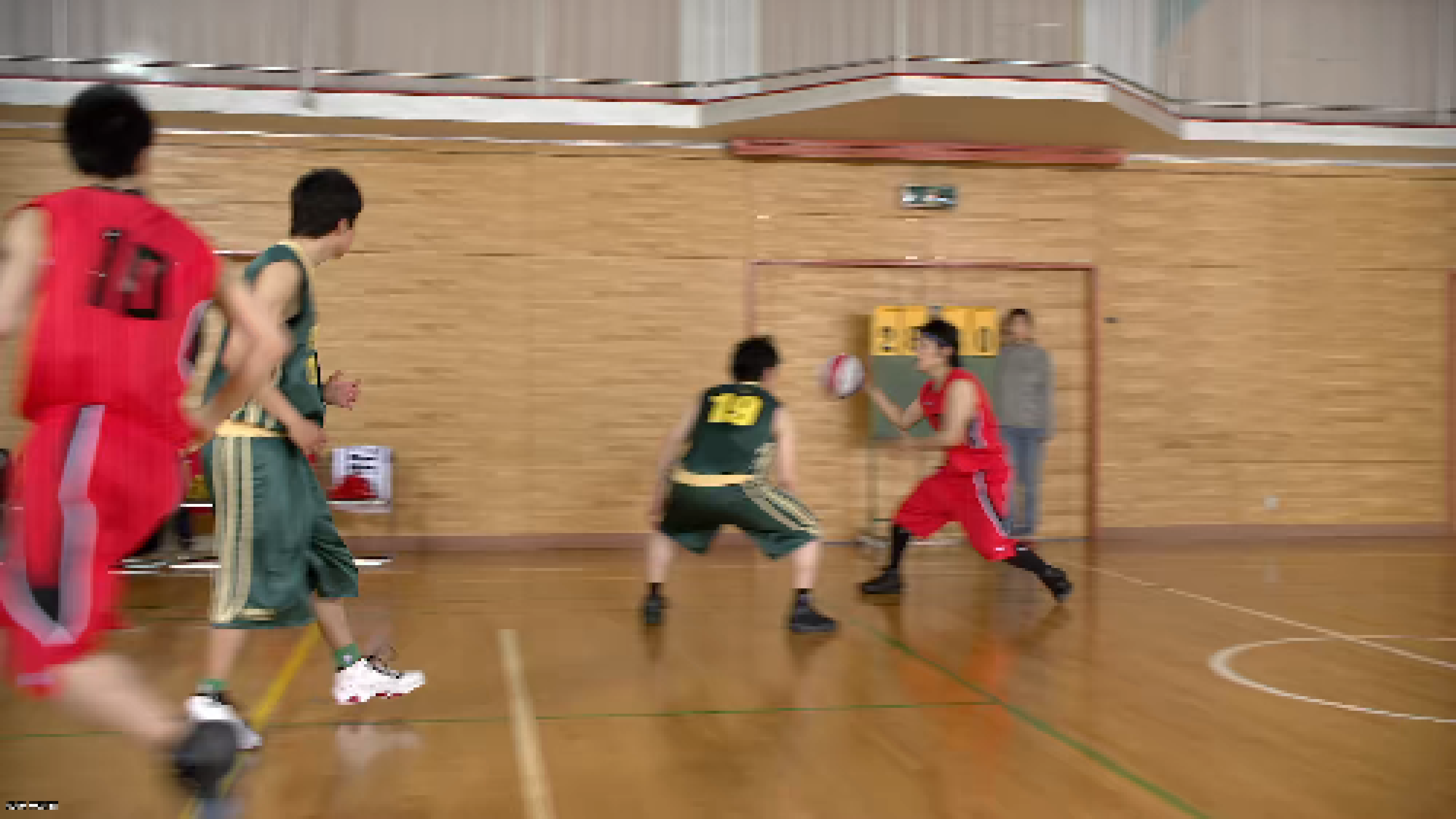}}
\subfigure[$\mbox{MSE-Y} = 10.8159$,
$\mbox{MSE-U} = 3.8290$,
$\mbox{MSE-V} = 3.5766$,
$\mbox{PSNR-Y} = 37.7902$,
$\mbox{PSNR-U} = 42.2999$,
$\mbox{PSNR-V} = 42.5961$]{\includegraphics[scale=.15]{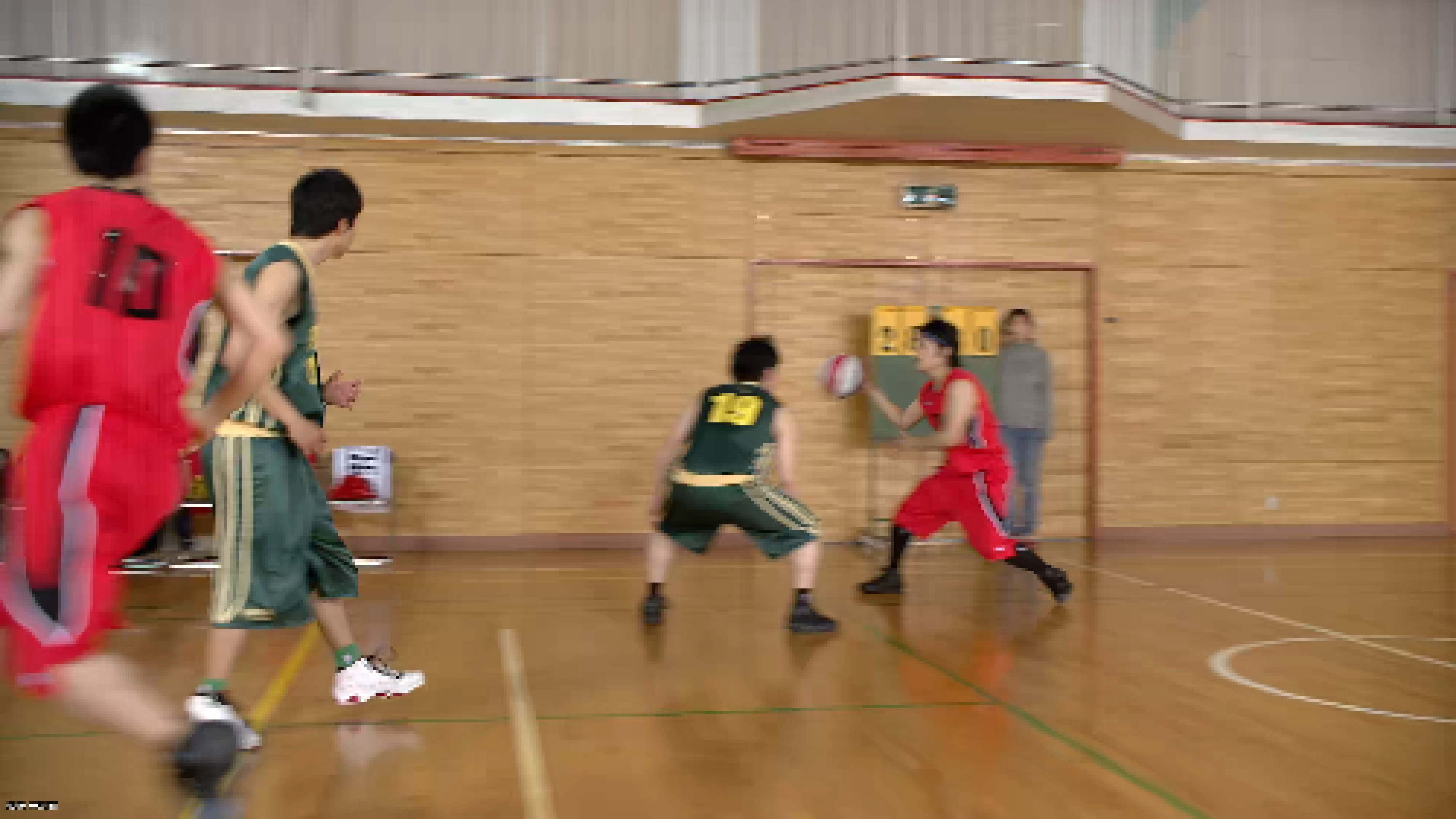}}

\subfigure[$\mbox{MSE-Y} = 10.1479$,
$\mbox{MSE-U} = 3.4765$,
$\mbox{MSE-V} = 3.1724$,
$\mbox{PSNR-Y} = 38.0670$, %
$\mbox{PSNR-U} = 42.7194$,
$\mbox{PSNR-V} = 43.1170$]{\includegraphics[scale=.15]{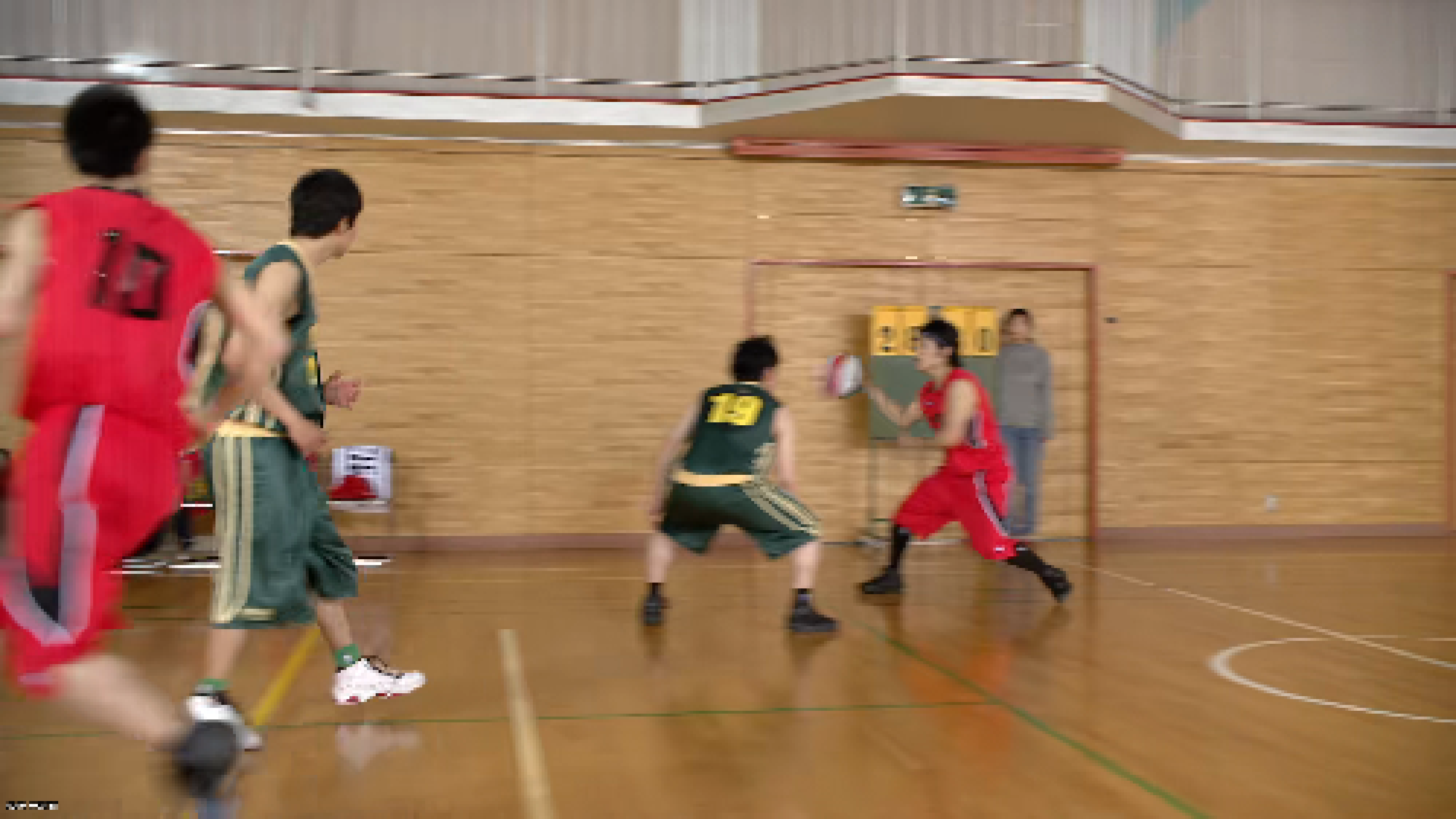}}
\subfigure[$\mbox{MSE-Y} = 10.3570$,
$\mbox{MSE-U} = 3.6228$,
$\mbox{MSE-V} = 3.3113$,
$\mbox{PSNR-Y} = 37.9785$,
$\mbox{PSNR-U} = 42.5403$,
$\mbox{PSNR-V} = 42.9308$]{\includegraphics[scale=.15]{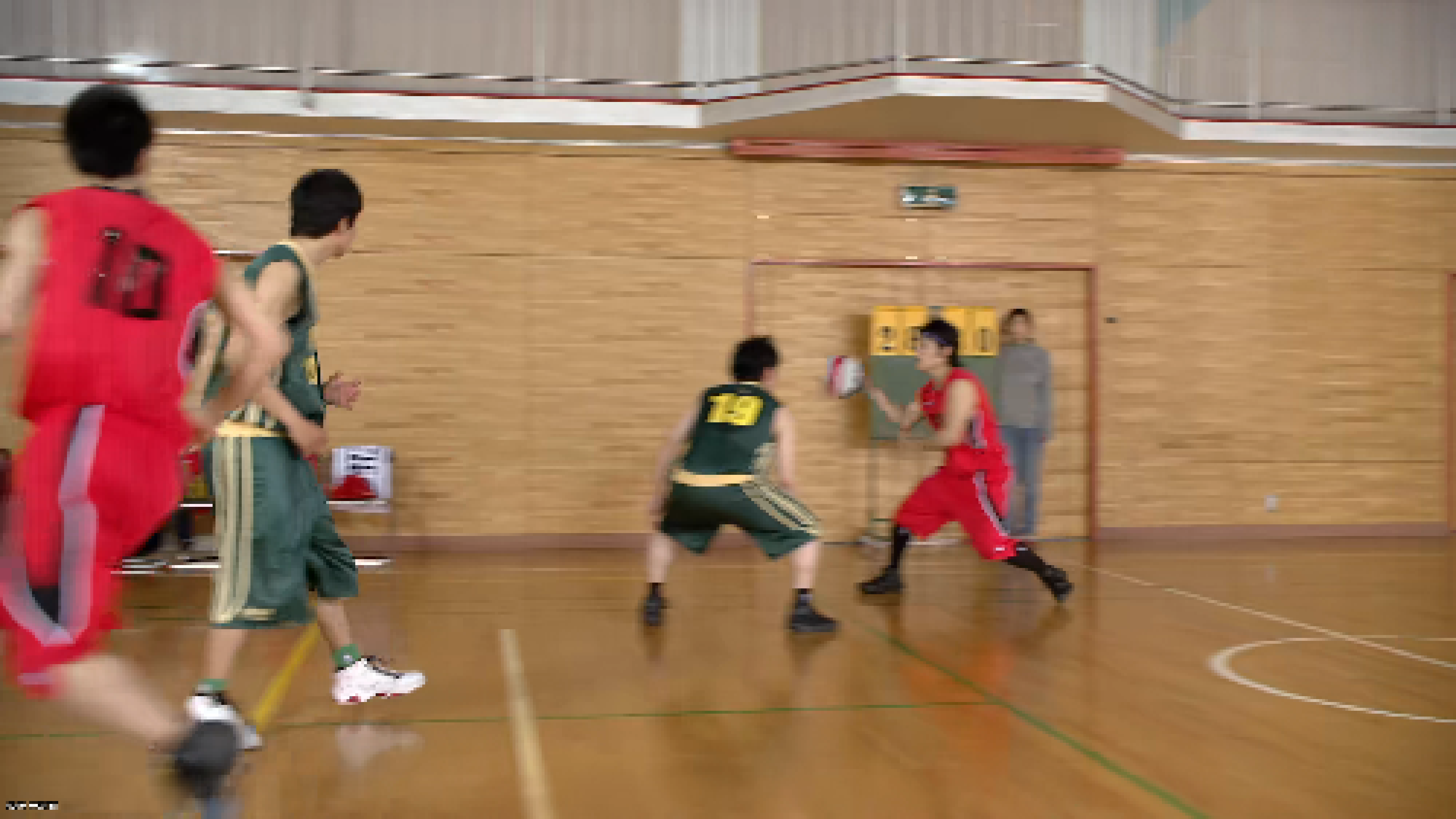}}

\subfigure[$\mbox{MSE-Y} = 14.0693$,
$\mbox{MSE-U} = 4.0741$,
$\mbox{MSE-V} = 4.4404$,
$\mbox{PSNR-Y} = 36.6481$,
$\mbox{PSNR-U} = 42.0304$,
$\mbox{PSNR-V} = 41.6566$]{\includegraphics[scale=.15]{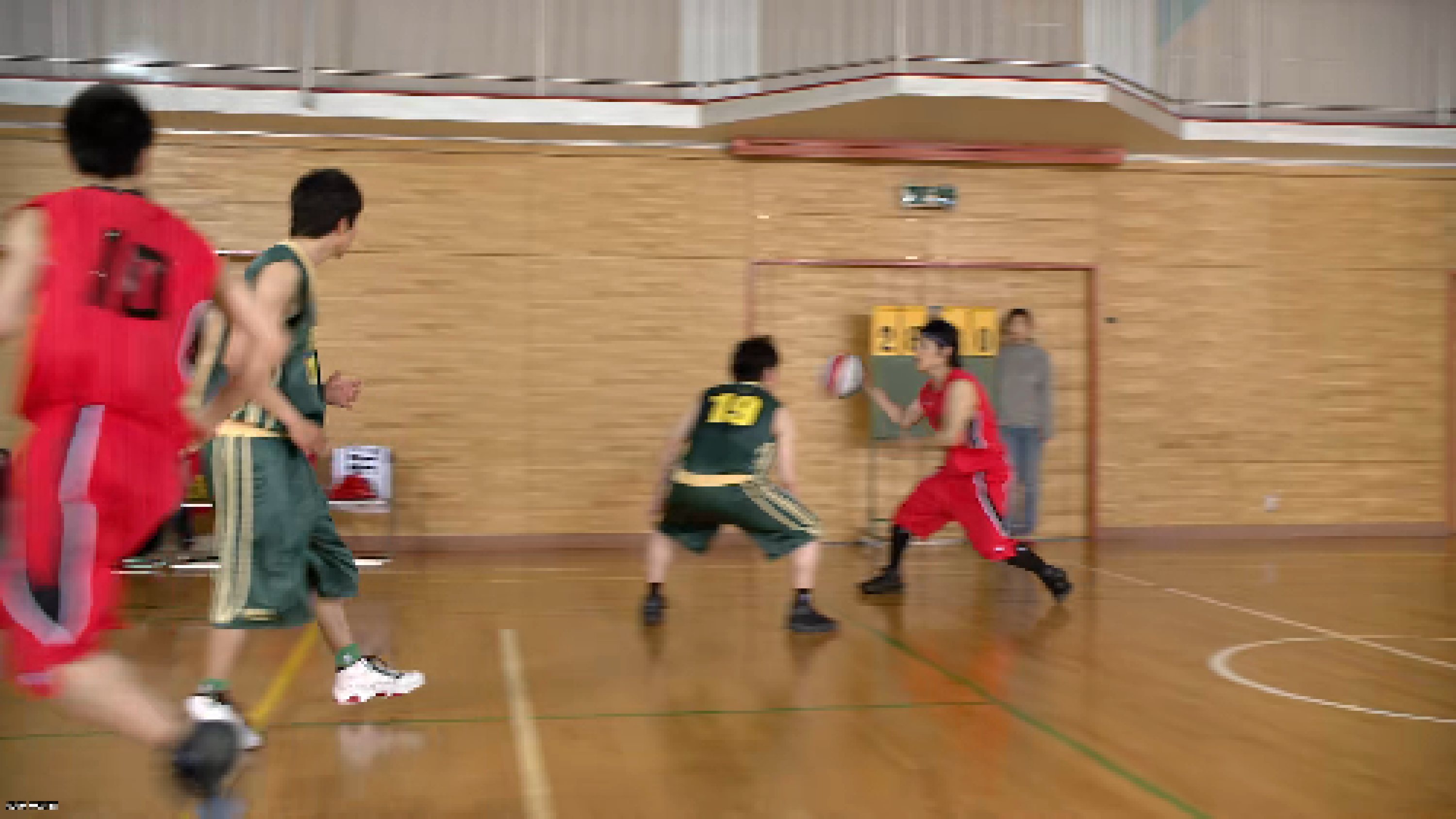}}
\subfigure[$\mbox{MSE-Y} = 14.5953$,
$\mbox{MSE-U} = 4.1377$,
$\mbox{MSE-V} = 4.6053$,
$\mbox{PSNR-Y} = 36.4887$,
$\mbox{PSNR-U} = 41.9632$,
$\mbox{PSNR-V} = 41.4982$]{\includegraphics[scale=.15]{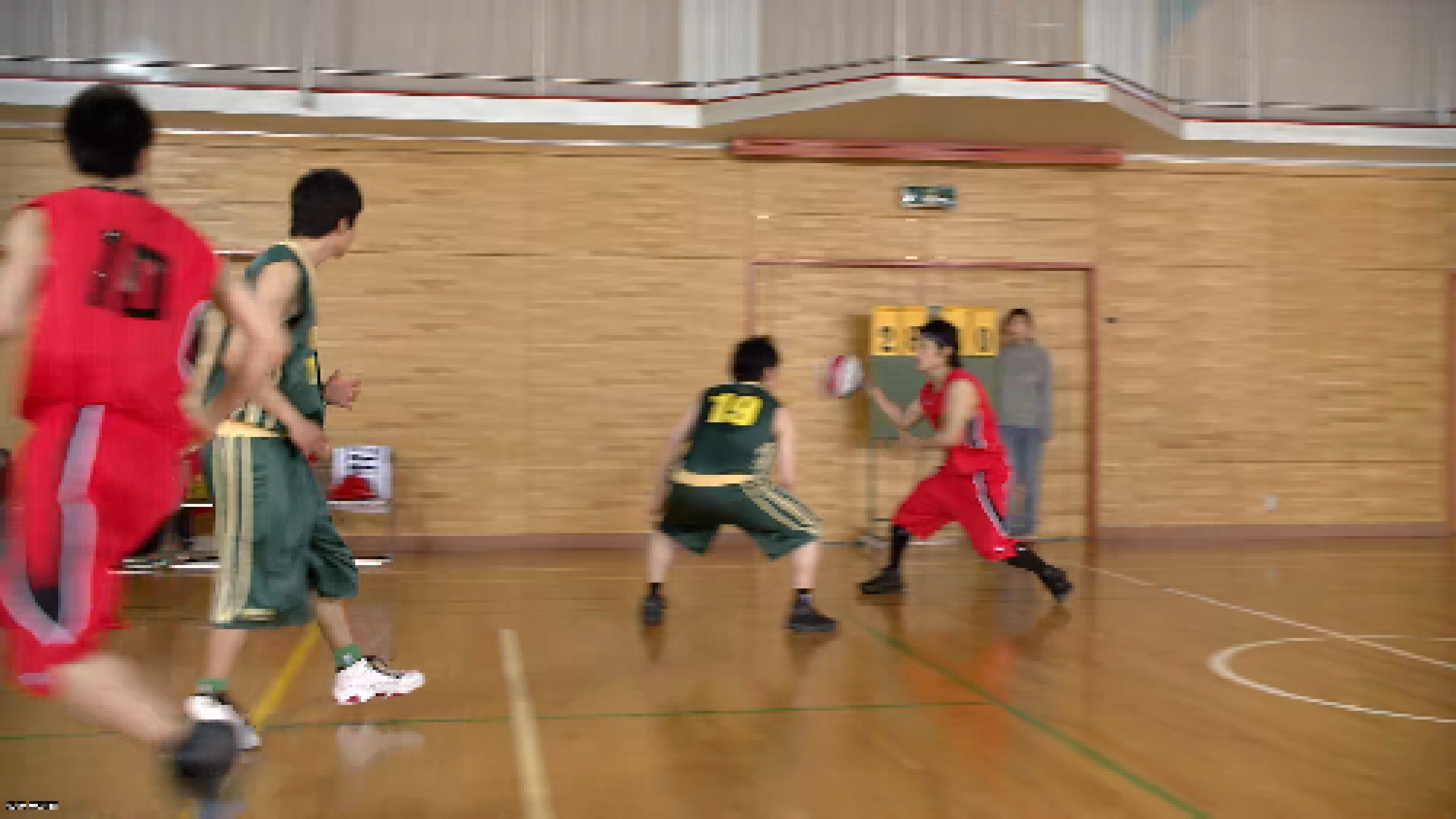}}

\subfigure[$\mbox{MSE-Y} = 14.6155$,
$\mbox{MSE-U} = 4.1349$,
$\mbox{MSE-V} = 4.5502$,
$\mbox{PSNR-Y} = 36.4827$,
$\mbox{PSNR-U} = 41.9661$,
$\mbox{PSNR-V} = 41.5505$]{\includegraphics[scale=.15]{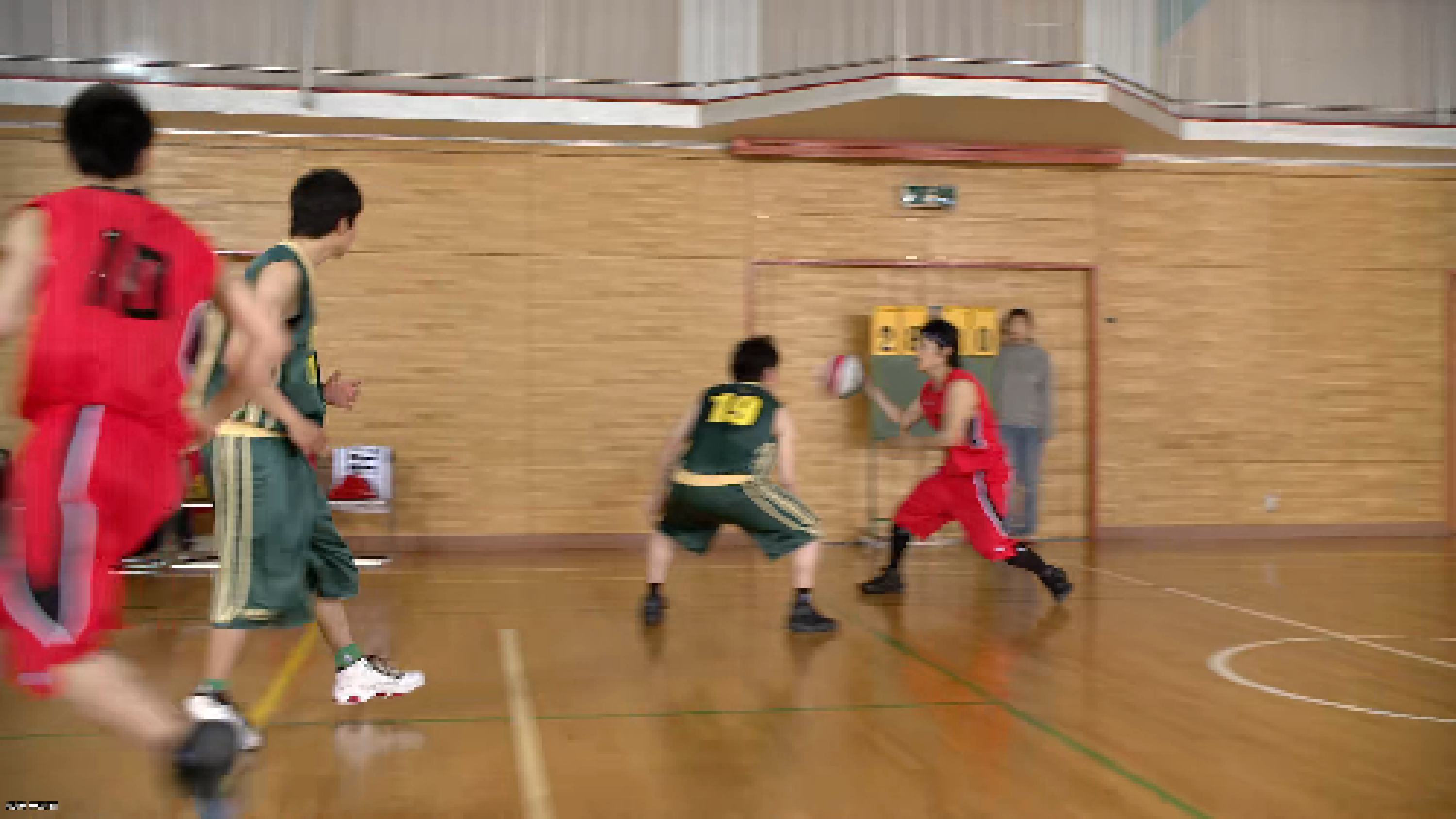}}
\subfigure[$\mbox{MSE-Y} = 15.0761$,
$\mbox{MSE-U} = 4.2812$,
$\mbox{MSE-V} = 4.6444$,
$\mbox{PSNR-Y} = 36.3479$,
$\mbox{PSNR-U} = 41.8151$,
$\mbox{PSNR-V} = 41.4615$]{\includegraphics[scale=.15]{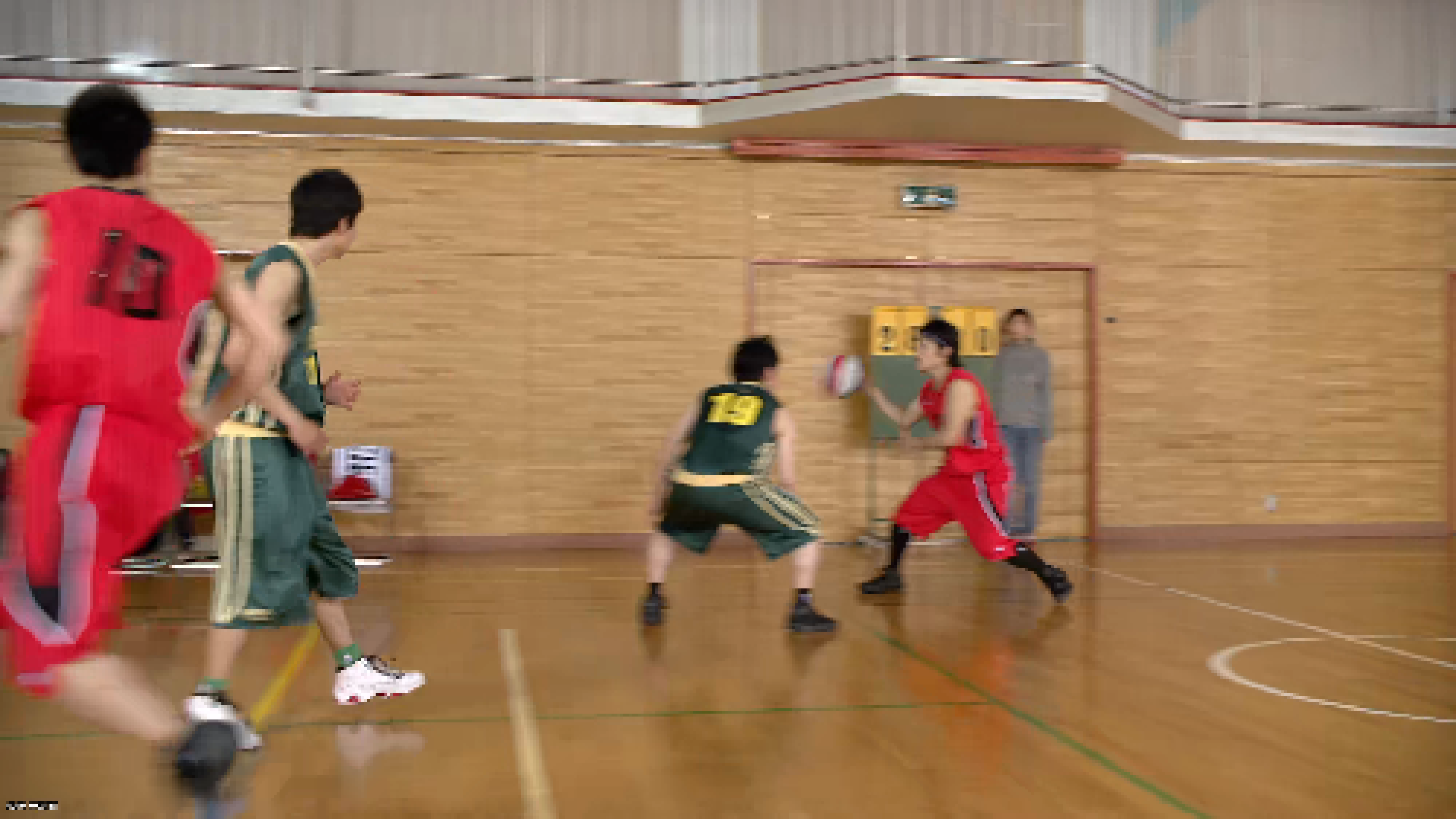}}
\caption{
Compression of the tenth frame of \texttt{BasketballDrive} using
(a),(c),(e)~the default and (b),(d),(f)~the modified versions of the HEVC software
for $\mbox{QP} = 32$, and AI, RA, LD-B, and LD-P coding configurations, respectively.}
\label{fig:exemplehevc}
\end{figure*}

Despite the very low computational cost
when compared to the IDCT (cf. Table~\ref{tab:hevccost}),
the proposed transform
does not introduce
significant errors.
Figure~\ref{fig:exemplehevc}
illustrates
the tenth frame of
the \texttt{BasketballDrive} video
encoded according to~the default HEVC IDCT
and~$\mathbf{\widehat{C}}_1$ and its scaled versions
for each coding configuration.
The QP was set to 32.
Visual degradations
are virtually nonperceptible
demonstrating
real-world applicability
of the proposed DCT approximations
for high resolution video coding.

\begin{table*}
\setlength{\tabcolsep}{3pt}
\centering
\small
\caption{
BD-PSNR (dB) and BD-Rate (\%) of the modified HEVC reference software for tested video sequences}
\label{tab:bdpsnrbdrate}
\begin{tabular}{@{}lcccccccc@{}}
\toprule
\multirow{2}{*}{Video sequence}  & \multicolumn{2}{c}{AI} & \multicolumn{2}{c}{RA} & \multicolumn{2}{c}{LD-B} & \multicolumn{2}{c}{LD-P} \\
& BD-PSNR & BD-Rate & BD-PSNR & BD-Rate & BD-PSNR & BD-Rate & BD-PSNR & BD-Rate \\
\midrule
\texttt{PeopleOnStreet}      & 0.2999 & $-5.5375$ & 0.1467& $-3.4323$&   N/A &     N/A   & N/A   & N/A\\
\texttt{BasketballDrive}     & 0.1692 & $-6.1033$ & 0.1412& $-6.1876$& 0.1272& $-5.2730$ & 0.1276& $-5.2407$\\
\texttt{RaceHorses}          & 0.4714 & $-5.8250$ & 0.5521& $-8.6149$& 0.5460& $-7.9067$ & 0.5344& $-7.6868-$ \\
\texttt{BlowingBubbles}      & 0.0839 & $-1.4715$ & 0.0821& $-2.1612$& 0.0806& $-2.1619$ & 0.0813& $-2.2370$ \\
\texttt{KristenAndSara}      & 0.2582 & $-5.0441$ & N/A   &      N/A & 0.1230& $-4.1823$ & 0.1118& $-4.0048$ \\
\texttt{BasketballDrillText} & 0.1036 & $-1.9721$ & 0.1372& $-3.2741$& 0.1748& $-4.3383$ & 0.1646& $-4.1509$ \\
\bottomrule
\end{tabular}
\end{table*}

\section{Conclusion}
\label{s:conclusion}

In this paper,
we set up and solved
an optimization problem
aiming at the proposition
of new approximations
for the 8-point DCT.
The obtained approximations
were determined
according to a greedy heuristic
which minimized the angle
between the rows of the approximate
and the exact DCT matrices.
Constraints of orthogonality
and low computational complexity
were imposed.
One of the obtained approximations
outperformed all the considered approximations
in literature
according to
popular performance measures.
We also introduced
the use of circular statistics
for assessing
approximate transforms.
For the proposed transform $\mathbf{T}_1$,
a fast algorithm
requiring only 24~additions and 6~bit-shifting operations
was proposed.
The fast algorithm for the proposed method
and directly competing
approximations
were given
FPGA realizations.
Simulations were made
and the hardware resource consumption and power consumption
were measured.
The maximum operating frequency
of the proposed method
was 37.4\% higher
when compared with
the well-known
Lengwehasatit--Ortega approximation~(LO)~\cite{ortega2004}.
In addition,
the applicability
of the proposed approximation
in the context of image compression
and video coding
was demonstrated.
Our experiments
also demonstrate
that DCT approximations
can effectively approximate the DCT behavior,
but also---under particular conditions---outperform
the DCT itself
for image coding.
The proposed approximation
is fully HEVC-compliant,
being
capable of video coding
with HEVC quality
at lower computational costs.

{\small
\singlespacing
\bibliographystyle{IEEEtran}
\bibliography{references}
}

\end{document}